\documentclass[compsoc,conference,a4paper,10pt,times]{IEEEtran}
\IEEEoverridecommandlockouts
\usepackage{cite}
\usepackage{amsmath,amssymb,amsfonts}
\usepackage{algorithmic}
\usepackage{graphicx}
\usepackage{textcomp}
\usepackage{bmpsize}
\usepackage{xcolor}
\usepackage{lipsum}
\usepackage[colorlinks=true,urlcolor=black]{hyperref}
\usepackage{array}
\usepackage{placeins}
\usepackage{multicol}
\usepackage{multirow}
\usepackage{fancyhdr}
\usepackage{dblfloatfix}
\usepackage{balance}

\usepackage[normalem]{ulem}
\usepackage{tikz}
\newcommand*\circled[1]{\tikz[baseline=(C.base)]{\node[circle, draw, inner sep=0.3pt, fill=black, text=white] (C) {\vphantom{1g}#1};}}

\def\BibTeX{{\rm B\kern-.05em{\sc i\kern-.025em b}\kern-.08em
    T\kern-.1667em\lower.7ex\hbox{E}\kern-.125emX}}

\makeatletter
\newcommand*\titleheader[1]{\gdef\@titleheader{#1}}
\AtBeginDocument{%
  \let\st@red@title\@title
  \def\@title{%
    \bgroup\normalfont\large\centering\@titleheader\par\egroup
    \vskip1.5em\st@red@title}
}
\makeatother

\begin{document}

\pagenumbering{arabic}
\setcounter{page}{1}

\title{The Avatar Cache: Enabling On-Demand Security with Morphable Cache Architecture}


\author{
    \IEEEauthorblockN{Anubhav Bhatla$^*$}
	\IEEEauthorblockA{MIT\\abhatla9@mit.edu}
	\and
	\IEEEauthorblockN{Navneet Navneet$^*$}
	\IEEEauthorblockA{Georgia Tech\\n41@gatech.edu}
	\and
	\IEEEauthorblockN{Moinuddin Qureshi}
	\IEEEauthorblockA{Georgia Tech\\moin@gatech.edu}
    \and
	\IEEEauthorblockN{Biswabandan Panda}
	\IEEEauthorblockA{IIT Bombay\\biswa@cse.iitb.ac.in}
}

\maketitle

\def\thefootnote{*}\footnotetext{Equal contribution}\def\thefootnote{\arabic{footnote}}

\begin{abstract}
The sharing of the last-level cache (LLC) among multiple cores makes it vulnerable to cross-core conflict and occupancy-based attacks. Even after significant research in secure LLC designs, modern processors employ non-secure set-associative LLCs. In general, there are two possible secure LLC designs: (i) a randomized LLC and (ii) a partitioned LLC.
The state-of-the-art randomized LLC design, Mirage, mitigates conflict-based attacks. However, it incurs significant area overhead (20\% additional storage) and design complexity, with marginal performance overhead. On the other hand, LLC partitioning techniques mitigate cache occupancy-based attacks as well as conflict-based attacks. However, partitioning techniques incur significant performance overheads (on average more than 5\% and as high as 49\%). Additionally, set-based partitioning requires OS support, whereas way-based partitioning techniques suffer from scalability issues. These factors pose major obstacles to the industrial adoption of secure LLCs. 
This paper investigates whether LLC security can be achieved with minimal modifications to a conventional set-associative LLC, enabling morphable secure LLC design (security only when needed) while keeping performance, power, and area overheads low.


We propose Avatar cache, a secure, morphable LLC with three modes—\textit{non-secure} (Avatar-N), \textit{randomized secure} (Avatar-R), and \textit{partitioned secure} (Avatar-P)—and the ability to switch dynamically between them. Its design closely resembles a conventional set-associative LLC, facilitating industry adoption.
Avatar-R provisions extra invalid entries to provide a strong security guarantee. By leveraging high associativity, it maintains these entries without significantly reducing cache capacity. Avatar can also morph into a partitioned LLC in Avatar-P to mitigate both conflict-based and occupancy-based attacks.
Avatar-R mitigates conflict-based attacks with a stronger security guarantee than Mirage—only one set-associative eviction per {\boldmath $10^{30}$} years—while incurring just 1.5\% storage overhead, a 2.7\% increase in static power consumption, and a mere 0.2\% slowdown compared to a non-secure 16 MB baseline LLC. Avatar-P has a slightly higher 3\% performance overhead compared to the non-secure baseline, performing significantly better than the state-of-the-art way-based partitioned LLC. When security is not a priority, we can switch to Avatar-N, morphing into a traditional LLC to optimize performance and power efficiency.



\end{abstract}


\section{Introduction}

The shared last-level cache (LLC) plays a crucial role in mitigating long-latency memory accesses and enhancing the system's overall performance. 
The LLC is susceptible to side-channel attacks, which can leak sensitive information. Previous works, such as Prime+Probe~\cite{prime+probe} and Evict+Reload~\cite{evict+reload}, have demonstrated that it is possible to create controlled contention in LLC sets and then measure the latency differences between an LLC hit and an LLC miss to observe the effects of this contention. In conventional set-associative caches, a set of addresses is mapped to a small region in the LLC. An efficient adversary can access the cache in a particular manner to cause set conflicts, where the victim evicts the attacker's lines and vice versa. As both memory accesses map to the same set, this is referred to as a set-associative eviction (SAE), and it is the primary enabler of the attacks mentioned above. 

A fully associative LLC mitigates conflict-based attacks. Mirage~\cite{mirage} and Maya~\cite{maya}, two state-of-the-art randomized LLC designs, mimic a fully associative LLC by decoupling insertion and eviction, making it immune to conflict-based side-channel attacks. 
These designs do not need OS support and incur marginal performance overheads. 
However, a new class of LLC attack, called occupancy-based attack~\cite{occupancy}, exploits cache occupancy to leak information, and most randomized LLC designs, including Mirage and Maya, fail to address this threat~\cite{usec2025}. Secure LLC partitioning techniques~\cite{sass, mi6, DAWG, BCE, teeshirt} are the only effective mitigation against occupancy-based attacks but incur a high performance overhead, as high as 49\% (for \texttt{cam4} on 8-core 16MB LLC with BCE~\cite{BCE}), and an average of at least 5\%. In addition, set-based partitioning techniques require OS support for fine-grained LLC partitions~\cite{mi6,BCE}. Way-based partitioning, conversely, needs minimal OS support, but its scalability is limited by cache associativity. SassCache~\cite{sass} is a randomization-based partitioning solution, providing strong security guarantees against both conflict-based and occupancy-based attacks. However, it suffers from high performance overheads (16\% average slowdown\footnote{We conducted in-house simulations to determine these results, as the original study did not evaluate the geomean slowdown.} on a 2-core system).


\vspace{0.05 in}
\noindent \textbf{The problem.} 
State-of-the-art randomized LLCs, such as Mirage and Maya, mitigate conflict-based attacks and introduce substantial design complexity compared to traditional non-secure set-associative LLCs. Their decoupled tag and data store necessitate pointers, further complicating implementation.  
On the other hand, secure LLC partitioning techniques offer strong isolation against all LLC attacks, albeit at the expense of performance and scalability.

\vspace{0.05 in}
\noindent \textbf{Pertinent question.} Why incur such high overheads and design complexity for applications that do not require security? Moreover, the high overheads and design complexity make secure LLC designs impractical for industry adoption as they diverge significantly from conventional set-associative LLCs.

\vspace{0.05 in}
\noindent \textbf{Our requirements.}
First, from a commercial deployment perspective, the industry demands a simple LLC design that is not significantly different from a traditional non-secure set-associative LLC in terms of design overhead and runtime overhead while providing robust security against LLC attacks. Recent proposals fail to meet this practical industry requirement. Second, we want a complete microarchitecture solution for security with minimal support from the OS, contrary to recent partitioned designs. Third, from the user’s perspective, a secure LLC is not required for all applications; thus, incurring significant performance and power overheads for non-sensitive workloads is inefficient.

\vspace{0.05 in}
\noindent \textbf{Our goal} is to propose a secure LLC design that offers \textit{security-on-demand}, i.e., one can choose security against LLC attacks only if required. If one chooses to enable security by operating in \textit{secure} mode, it should come with low design complexity and runtime overhead, making industry adoption easier. At the same time, if security is not required, the same LLC design should seamlessly operate in a \textit{non-secure} mode, morphing into a traditional set-associative LLC.


\begin{figure}[t]
    \centering
    \includegraphics[width=0.94\linewidth]{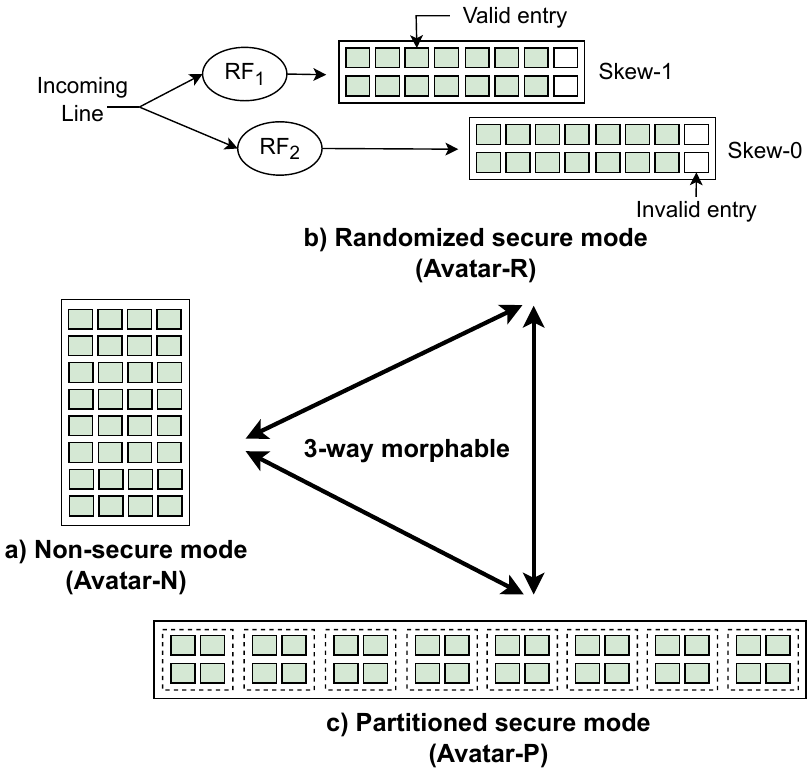}
    \caption{Overview of the Avatar LLC design with all its operating modes. RF is the randomizing function implemented using a block cipher.}
    \label{fig:overview}
\end{figure}

\vspace{0.05 in}
\noindent \textbf{Our approach.} We observe that increasing LLC associativity can help us design a state-of-the-art randomized LLC like Mirage without any design-time overhead. Similarly, a highly associative LLC can facilitate a scalable and high-performing way-partitioned LLC. To enable on-demand security, we offer configurable security parameters—one ensuring protection against all attacks with high runtime overhead and another providing limited protection with lower overhead.  

We propose the Avatar Cache, a morphable LLC substrate, which is a practical and secure LLC design with the ability to morph between secure and non-secure modes during runtime, providing \textit{security-on-demand}.
Figure~\ref{fig:overview} provides an overview of the Avatar cache. Avatar cache operates in three modes:
(i) \textit{non-secure} mode (\textbf{Avatar-N}), (ii) \textit{randomized secure} mode (\textbf{Avatar-R}), and (iii) \textit{partitioned secure} mode (\textbf{Avatar-P}). In Avatar-R, to mitigate conflict-based attacks, the Avatar cache morphs into a randomized skewed-associative LLC (the LLC is split into two halves) with high associativity, providing a stronger security than Mirage and Maya, while incurring ultra-low storage, performance, and power overheads. To mitigate occupancy-based attacks, the user can transform the cache into a way-based partitioned LLC with high associativity in the Avatar-P mode. Avatar can also operate as a traditional non-secure set-associative LLC in the Avatar-N mode based on the user's security requirements. This morphability is achieved by designating Model Specific Register (MSR) bits for the operation mode, which have previously been used to configure various microarchitecture features (e.g., hardware prefetchers~\cite {msr}).


\vspace{0.05 in}
\noindent \textbf{The key idea} of Avatar-R's defense against conflict-based attacks is to use a skewed randomized LLC design and provision extra invalid ways. Each skew is associated with a cipher for address-to-set mapping randomization with a fraction of the cache entries (both tag and data store) \emph{invalid} at all times with \emph{implicit} indirection instead of a Mirage-like explicit indirection to provision extra invalid tag ways. In Avatar-R, the LLC operates at high associativity with two skews, which helps reduce the cache capacity attributed to extra invalid ways, thus improving the usable cache capacity and providing a sweet spot in terms of security guarantees and design and runtime overheads. When it comes to mitigating occupancy-based attacks, the Avatar cache can morph into Avatar-P, which facilitates a way-based partitioned LLC with high associativity but without skews. 

In summary, for an 8-core system, we transform a 16 MB 16-way LLC into Avatar-R with 128 ways per skew (total 256 ways between two skews) or into Avatar-P with 256 ways (and no skews), allowing partitioning across up to 256 security domains. We ensure that every mode switch comes with a complete LLC flush, and we also ensure not to enable new LLC attacks, including denial-of-service (DoS) attacks or flush latency-based attacks.

\vspace{0.05 in}
\noindent \textbf{Our contributions.} \\
\noindent(i) We present Avatar, a morphable LLC substrate that enables dynamic switching between \textit{non-secure} (Avatar-N), \textit{randomized secure} (Avatar-R), and \textit{partitioned secure} (Avatar-P) modes. This allows the LLC to operate either as a secure cache or as a traditional set-associative cache with minimal changes (Section~\ref{sec:design}). \\
(ii) Avatar-R re-imagines Mirage using high associativity without incurring the 20\% storage overhead by employing implicit tag-to-data indirection. Avatar-P enables scalable, high-associativity way-based partitioning to defend against occupancy-based attacks (Sections~\ref{sec:design} and~\ref{sec:security}). \\
(iii) We provide a framework for dynamic mode switching that ensures low runtime overhead and prevents new attack vectors, including denial-of-service (DoS) attacks and flush latency-based attacks. We show that it reduces the design and runtime overheads, making complete LLC security on-demand and yet practical (Section~\ref{sec:switch}). We also quantify the performance, power, and area overheads (Section~\ref{sec:evaluation} and~\ref{sec:Overheads}).  

\section{Background}

\subsection{Threat model}
\label{sec:threat}

We assume the attacker and victim run on separate cores in a multi-core system, sharing the LLC. Private caches are spatially and temporally partitioned among processes and flushed on context switches for isolation. The attacker can access the LLC by sending memory requests to their own data but cannot access cache lines outside their address space. However, the attacker can accurately distinguish LLC hits from misses in their data accesses. The attacker can flush their cache blocks to expedite eviction set creation and employ various strategies to do so. In the case of the randomized secure mode, the attacker possesses full knowledge of the randomized cache design and encryption algorithm but lacks access to the block cipher key used for set mapping.
\subsection{LLC contention attacks}
\label{sec:contention}

\noindent \textbf{Eviction-based attacks.} The attacker \textit{primes} their data into the LLC such that they cause conflicts with the victim's data and cause an SAE. Later, in the \textit{probe} step, the attacker re-accesses their data and uses the timing information to infer the victim's memory access pattern. If the attacker observes a longer latency, it implies that the victim has evicted some of the attacker's previously inserted cache lines~\cite{prime+probe}. It is also known as a conflict-based attack. A recent attack~\cite{seed2022} showed a variation of this using probabilistic eviction sets.

\noindent \textbf{Shared-memory based attacks.} The attacker leverages the fact that they share their address space with the victim (e.g., shared libraries). For example, in the Flush+Reload~\cite{flush+reload} attack, the attacker flushes these cache lines out of the LLC, and later observes the victim's access to these shared cache lines by using the timing difference in memory access latency.


\noindent \textbf{Flush-based eviction attacks.} The attacker can mount an eviction-based attack by flushing their private data while creating an eviction set~\cite{fixrandomized}. This method of attack is faster than conventional eviction-based attacks like Prime+Probe. This attack type differs from shared memory-based flush attacks, where the attacker flushes shared cache lines from the LLC. 

\noindent \textbf{Occupancy-based attacks.} The attacker observes the LLC space occupied by the victim and manipulates their occupancy to induce cache conflicts or evictions, leading to side-channel leaks. Recent attacks like website fingerprinting~\cite{occupancy} exploit the dynamic LLC usage between the attacker and the victim.


\subsection{Secure randomized LLCs}
\label{sec:randomization}
Randomized LLCs provide the best performance and security tradeoff when it comes to mitigating conflict-based attacks. 
\noindent \textbf{CEASER~\cite{CEASER}} encrypts a physical address using a key to obtain an encrypted address during an LLC access. Despite the encryption, LLC conflicts can still occur, allowing an eviction-based attacker to launch an attack potentially. 
\noindent \textbf{CEASER-S~\cite{ceaser-s}} and \textbf{ScatterCache~\cite{scattercache}} advance beyond CEASER by proposing randomization with a skewed associative LLC to counter agile eviction-based LLC attackers that can exploit CEASER's slow remapping rates. CEASER, CEASER-S, and ScatterCache are simple designs, but their security guarantees are broken fast. 
\noindent \textbf{Mirage~\cite{mirage}} and \textbf{Maya~\cite{maya}} propose a fully associative LLC using multi-index randomization and global random eviction. They decouple the tag and data stores, with pointer-based indirection allowing set-associative tag lookup and random eviction in the data store. The skewed associative tag store utilizes additional invalid tags to prevent set conflicts, employing a load-aware policy to ensure the availability of these invalid tags. Mirage offers strong security, with only one SAE in $10^{17}$ years, but comes with a 20\% storage overhead, 17.5\% power overhead, and significant design complexity, requiring major changes to a traditional set-associative cache. Maya improves the storage and power overheads. 
However, the design complexity is similar to Mirage.

\subsection{Secure LLC partitioning techniques}
\label{sec:partitioning}

\noindent \textbf{Page coloring~\cite{mi6}} based LLC partitioning reduces cache conflicts by creating isolated regions at the LLC set level. 
However, this technique has limitations, as it cannot independently manage LLC and DRAM space. \textbf{Dynamically Allocated Way Guard (DAWG)~\cite{DAWG}} employs a software-configurable mask to manage way allocations among multiple security domains on a multi-core system. 
However, the available LLC ways limit DAWG's maximum number of isolated domains. For example, DAWG cannot be implemented on a 32-core system with a 16-way cache.
\noindent \textbf{Bespoke Cache Enclave (BCE)~\cite{BCE}} provides flexible cache partitioning to enhance data security, creating isolated cache partitions as small as 64 KB to protect against cache-based side-channel attacks. However, the partitions need to be resized and remapped after every allocation.
A similar principle as BCE is used in \textbf{Chunked-Cache~\cite{chunked-cache}} and \textbf{Composable Cachelets~\cite{composable-cachelets}}, where Chunked-Cache allows each program to have its dedicated cache sets. Composable Cachelets uses both way and set-based partitioning to provide support for dynamic reconfiguration of existing partitions. 
\noindent \textbf{SassCache~\cite{sass}} is a skewed associative cache with randomization that offers a soft-partitioning solution using cryptographic functions. However, it suffers from a significant increase in miss rate for memory-intensive GAP homogeneous workloads. According to our experiments, this results in a 16\% slowdown on a 2-core system. SassCache also cannot guarantee security against multiple adversary processes, which makes it vulnerable to new potential threat models, as highlighted in~\cite{sok}.
\textbf{INTERFACE~\cite{interface}} is a secure cache design which combines the skewed associative design of Mirage with partitioning to defend against conflict- and occupancy-based attacks. However, even after removing reverse pointers from data entries, they incur a heavy 11.62\% storage overhead and a 16.94\% power overhead, which makes it impractical for industry adoption.
\textbf{Ceviche~\cite{ceviche}} achieves fine-grained partitioning of the cache using a capability-based cache lookup. However, Ceviche has a performance slowdown of more than ~20\% for memory-bound workloads on an 8-core system. This places both SassCache and Ceviche on the same level as LLC partitioning techniques, which have similar performance overheads. 
\textbf{TEE-SHirT~\cite{teeshirt}} provides cache security within Trusted Execution Environments (TEEs). TEE-SHirT mitigates cache side-channel leakage by restructuring the cache hierarchy to block information leaks while ensuring scalability. It enhances security through a combination of set and way partitioning. \textbf{SCALE~\cite{scale}} employs dynamic partitioning by implementing bank-level way partitioning, where the number of partitions is constrained by the cache associativity multiplied by the number of banks. Dynamic cache-partitioning techniques improve cache utilization but at the cost of an occupancy attack. In general, static partitioning techniques completely mitigate occupancy-based attacks. Additionally, way-based partitioning is easy to implement and requires minimal OS support.

{
\begin{table}[htb]
  \caption{Utility of randomized and partitioned LLCs.}
    \centering
    \renewcommand{\arraystretch}{1.5}
    \resizebox{!}{1.9cm}{
    \begin{tabular}{|c||c|c|}
    \hline
        \textbf{Design} & \textbf{Randomized} & \textbf{Partitioned} \\ \hline  \hline
        \multicolumn{1}{|c||}{\multirow{2}{*}{\begin{tabular}[c]{@{}c@{}}\textbf{Security}\end{tabular}}} & \multicolumn{1}{c|}{\multirow{2}{*}{\begin{tabular}[c]{@{}c@{}}Conflict ($\checkmark$)\\[-0.5em]Occupancy ($\times$)\end{tabular}}} & \multicolumn{1}{c|}{\multirow{2}{*}{\begin{tabular}[c]{@{}c@{}}Conflict ($\checkmark$)\\[-0.5em]Occupancy ($\checkmark$)\end{tabular}}} \\
        \multicolumn{1}{|c||}{} & \multicolumn{1}{c|}{} & \multicolumn{1}{c|}{} \\ \hline
        \textbf{Scalability} & High & Low \\ \hline 
        \textbf{OS support} & No & Yes \\ \hline
        \textbf{Design time overhead} & High & Low \\ \hline
        \textbf{Runtime overhead} & Low & High \\ \hline

    \end{tabular}
    }
    \label{tab:summary}
\end{table}
}

\subsection{Goal: Practical and secure cache design}
As per Table~\ref{tab:summary}, both randomized and partitioned secure LLC architectures exhibit fundamental limitations. Randomized LLCs like Mirage and Maya provide strong security against conflict-based attacks but fail to protect against occupancy-based side-channel attacks. On top of this, they also suffer from high design complexity. Conversely, partitioned LLCs, while resilient to occupancy-based attacks, encounter scalability challenges, dependence on OS support, and incur considerable performance overheads due to suboptimal cache utilization. These trade-offs highlight the inadequacy of adopting a singular security mechanism across diverse workloads and threat models. Motivated by this, we propose Avatar—a unified, morphable LLC design that enables security to be provisioned on demand, in accordance with system or user requirements.

\section{Avatar cache: Design and implementation}
\label{sec:design}

\begin{figure}[htb]
    \centering
    \includegraphics[width=0.98\linewidth]{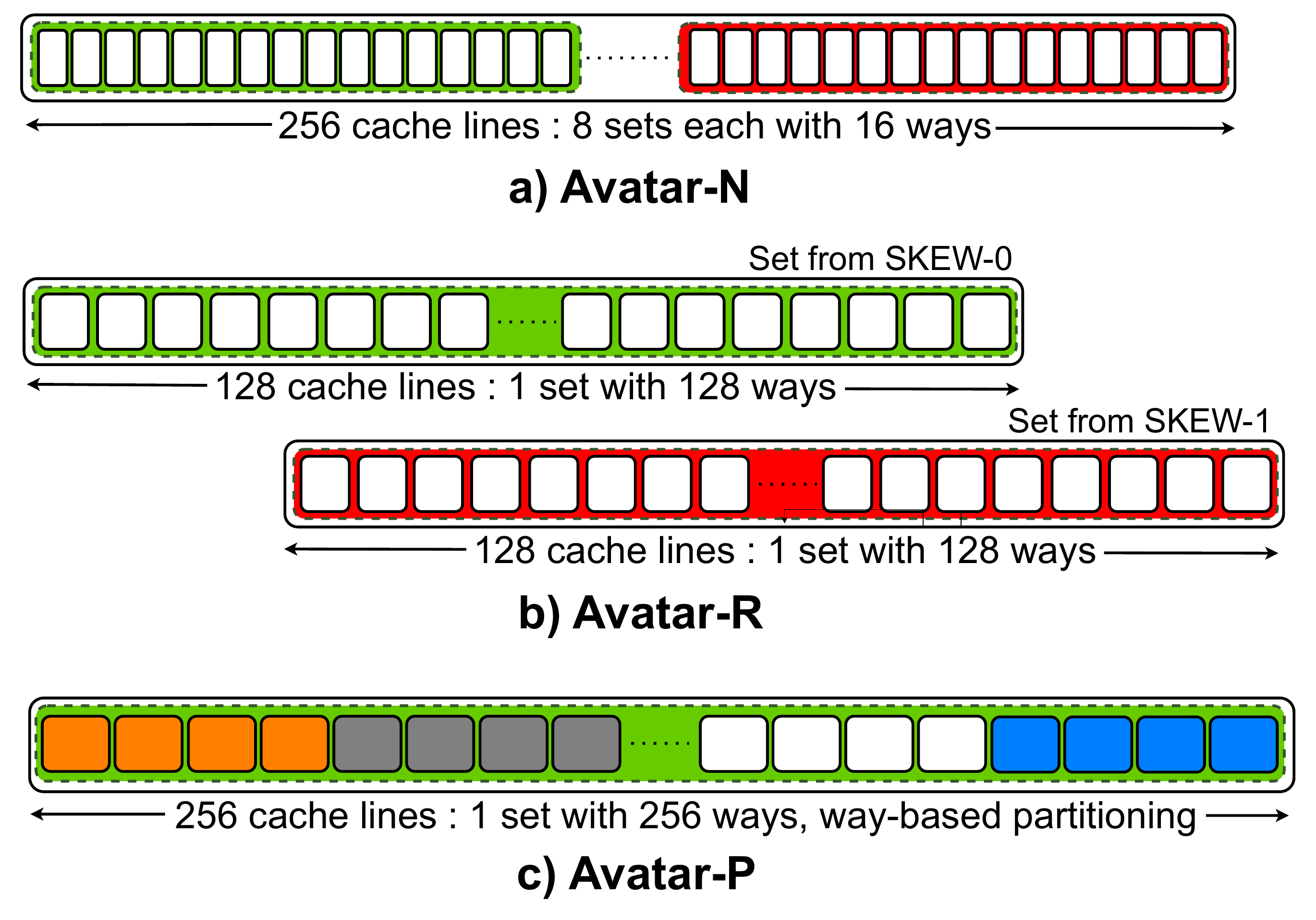}
    \caption{The morphing of sets and ways between the different modes in Avatar.
    }
    \label{fig:oasis_morph}
\end{figure}

The Avatar cache can operate in one of three modes: \textit{non-secure} (Avatar-N), \textit{randomized secure} (Avatar-R) or \textit{partitioned secure} (Avatar-P). In Avatar-N, the cache operates similarly to a traditional non-secure set-associative LLC. To protect the system against conflict-based attacks, Avatar can switch to Avatar-R, where the cache operates as a skewed randomized cache with high associativity and global random eviction as the replacement policy. 
Additionally, the Avatar cache can also be used as a partitioned cache with high associativity to provide much stronger security. In this Avatar-P mode, Avatar utilizes the existing high-associativity hardware and bypasses the cipher, as there is no need for randomization; instead, it provides a statically partitioned LLC. Figure \ref{fig:oasis_morph} shows how the LLC organization changes with each Avatar mode.

Avatar's two secure modes address different attacker threat models, offering flexibility to the user or security monitor~\cite{mi6, sanctum, Cyclone, CIPS} to select desired security-performance trade-offs based on workload requirements.

\subsection{Modes of operation}

\subsubsection{Avatar-N}
\label{sec:non-secure}
Avatar-N is intended for scenarios where security is not required and operates like a conventional set-associative LLC without skews. It employs smart replacement policies like Hawkeye~\cite{Jain2017HawkeyeL} to optimize performance and uses lower associativity than the secure modes to reduce power consumption.

\subsubsection{Avatar-R}
\label{sec:randomized}
Avatar-R mitigates SAEs by using a Mirage-like design and high associativity to provide security with minimal runtime overheads.

\vspace{0.05 in}
\noindent \textbf{Mirage with implicit mapping.} Avatar-R adopts a skewed associative randomized LLC architecture inspired by Mirage~\cite{mirage}, utilizing ciphers for randomized address-to-set mapping and a load-aware skew-selection policy~\cite{mirage} to evenly distribute cache lines across skews. The load-aware policy ensures that each incoming cache line can be filled in one of the two skew sets, depending on which set has more invalid entries. This means that both sets will need to be filled up completely to cause an SAE (we show that this is extremely improbable in Appendix~\ref{appendix:B}.

Unlike Mirage, which uses explicit indirection via pointers, Avatar-R employs implicit indirection between tag and data entries, significantly reducing overhead and enabling a lightweight, secure design. To preserve security, Avatar-R reserves a portion of the cache as invalid entries and operates at partial capacity. It also employs a \textit{global random eviction} policy, where a candidate is randomly selected from all valid LLC entries to prevent information leakage. A global counter tracks the number of valid LLC entries. Once it reaches a predefined threshold—corresponding to the maximum number of valid entries allowed to maintain the security guarantee—each subsequent fill triggers a global random eviction, where a random entry is evicted. If the counter is below the threshold, new lines are inserted directly without eviction. The seven invalid ways per skew are not enforced in every set; instead, this number represents the average number of invalid entries maintained across the entire cache through global random eviction.


\vspace{0.05 in}
\noindent \textbf{High associativity.} For Avatar-R, invalidating useful entries in a low-associativity cache can result in a significant performance loss. For example, in a 16-way associative cache (eight ways per skew), invalidating six ways per skew (this is important for providing complete security in the system's lifetime) leaves only 25\% of the cache capacity usable, drastically affecting performance. In contrast, a 256-way associative cache (128 ways per skew) would only require less than 5\% of its cache lines to be invalidated, making it a more practical solution. Thus, to maintain usable cache capacity while providing invalid ways, the Avatar LLC operates at high associativity, minimizing performance degradation. A recent study~\cite{sok} also highlighted the benefits of using high associativity to improve LLC security.


\subsubsection{Avatar-P}
\label{sec:partitioned}
Avatar-P mitigates both conflict- and occupancy-based attacks by employing way-based partitioning, wherein a fixed subset of ways is allocated to each security domain. It operates at high associativity, which not only reduces the performance overhead typically associated with LLC partitioning but also enables support for a significantly larger number of concurrent security domains compared to state-of-the-art partitioned LLCs~\cite{DAWG}. Like Avatar-N, it bypasses ciphers and employs a deterministic address-to-set mapping and leverages a smart replacement policy to enhance performance.

\begin{figure}[htp]
    \centering
    \includegraphics[width=\linewidth]{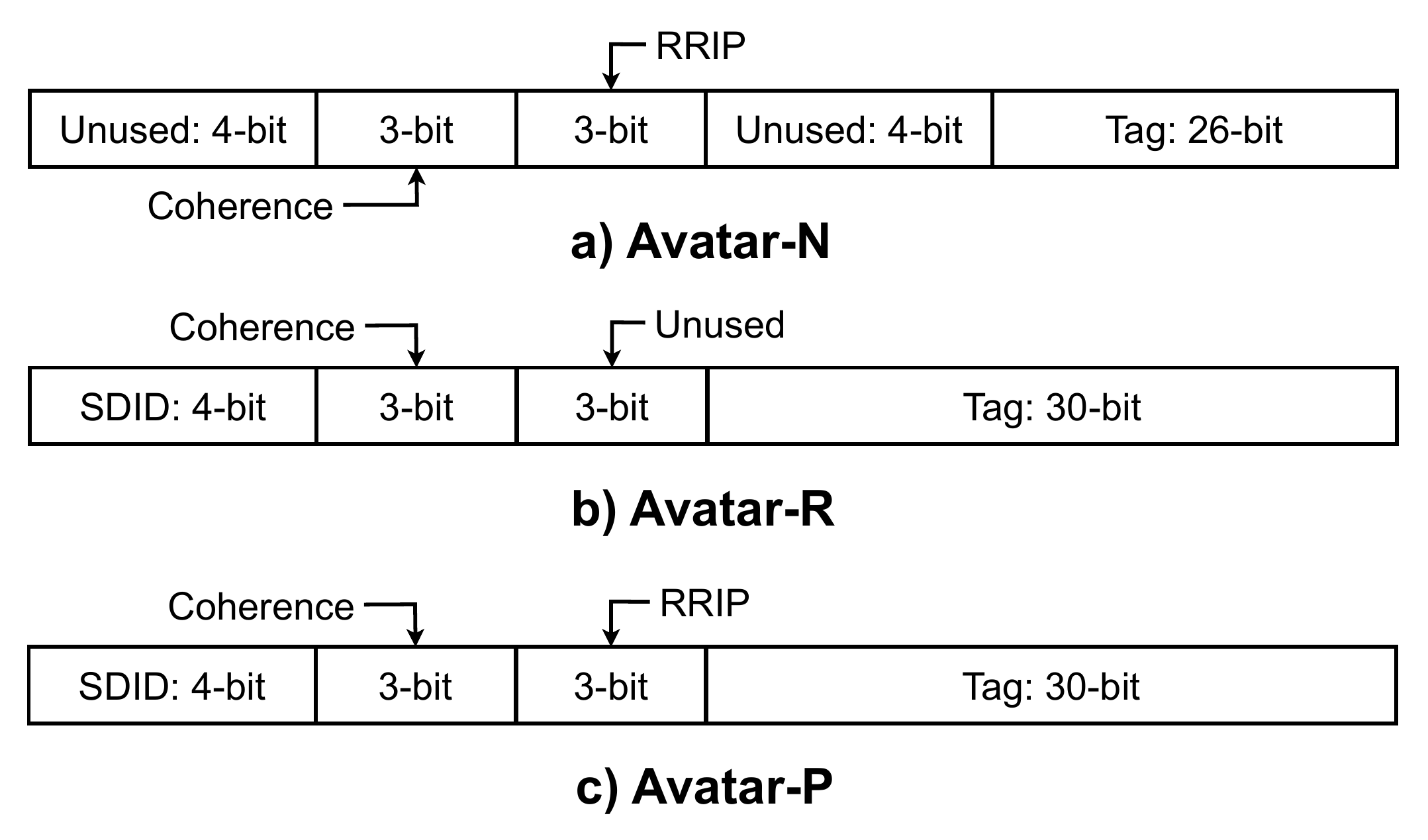}
    \caption{Tag bits in each of the possible modes of operation for Avatar. SDID is the secure domain ID and RRIP are the re-reference interval prediction~\cite{Jain2017HawkeyeL} replacement policy priority bits.
    }
    \label{fig:oasis_tag_entry}
\end{figure}

\subsection{Morphable Implementation}
\label{sec:implementation}

\subsubsection{Avatar-N}
For a 16 MB LLC, Avatar-N utilizes 16K sets with 16 ways per set and does not employ skews, requiring only 26 bits of the 46-bit line address for tag bits. Additionally, three coherence bits for the MOESI protocol and three re-reference interval prediction (RRIP) bits for the Hawkeye replacement policy~\cite{Jain2017HawkeyeL} are necessary. Figure \ref{fig:oasis_tag_entry} illustrates the organization of a tag entry in different operating modes of Avatar. The Avatar design allows the same hardware to be adapted for both modes. For a 16 MB LLC, Avatar-P uses all 40 bits to store useful information, while Avatar-N only utilizes 32 bits, leaving 8 bits unused per tag entry.

{
\begin{table} [htb]
  \caption{Overheads for the energy required per LLC access and performance as the associativity of Avatar-R varies from 128 ways to 512 ways, compared to a 16-way non-secure cache. Note that the LLC size is kept constant at 16 MB}.
    \centering
    \renewcommand{\arraystretch}{1.5}
   \resizebox{!}{0.95cm}{
    \begin{tabular}{|c||c|c|c|}
    \hline
        \textbf{Associativity} & \textbf{128-ways} & \textbf{256-ways} & \textbf{512-ways} \\ \hline \hline
        \textbf{Access Energy} & +3.5\% & +10\% & +25\% \\ \hline
        \textbf{Performance} & -0.9\% & -0.6\% & -0.5\% \\
    \hline
    \end{tabular}
    }
    \label{tab:performance-ass}
\end{table}
}

\subsubsection{Avatar-R}
Avatar-R utilizes two skews and leaves some cache entries invalid to prevent SAEs. Table \ref{tab:performance-ass} presents the per-access energy and performance overheads (based on SPEC CPU2017 workloads~\cite{spec}) for an 8-core system using a 16 MB Avatar cache, with associativity varying from 128 ways (64 ways per skew) to 512 ways (256 ways per skew) and seven invalid ways per skew. As associativity increases, a larger percentage of the cache becomes usable; for instance, with 128 ways, only 89\% of the cache is usable, increasing to 95\% with 256 ways and 97\% with 512 ways. Consequently, the 512-way configuration exhibits the least performance overhead. However, higher associativity also increases energy consumption, as more lines must be accessed for tag matching with each cache access. The 256-way configuration strikes a balance, offering performance nearly equivalent to the 512-way setup but with significantly lower energy requirements.

\begin{table}[htb]
  \caption{LLC line installs per SAE as the invalid ways vary from 5 to 7 ways per skew for Avatar-R with 256 ways (128 ways per skew).}
    \centering
    \renewcommand{\arraystretch}{1.5}
   \resizebox{!}{1.2cm}{
    \begin{tabular}{|c||c|}
    \hline
        \textbf{Invalid ways per skew} & \textbf{Installs} \\ \hline \hline
        5 invalid ways per skew & $2 \! \cdot \! 10^{12}$ (40 mins) \\ \hline
        6 invalid ways per skew & $10^{23}$ ($3 \! \cdot \! 10^{6}$ yrs) \\
        \hline
        7 invalid ways per skew & $5 \! \cdot \! 10^{46}$ ($10^{30}$ yrs) \\ \hline
    \end{tabular}
    }
    \label{tab:invalid-ways}
\end{table}

Table \ref{tab:invalid-ways} quantifies the security guarantee of Avatar-R in terms of the number of cache lines needed to trigger an SAE with varying invalid ways per skew (refer to Section \ref{sec:security} for more details). With seven invalid ways per skew, Avatar-R provides complete protection against eviction-based attacks throughout the system's lifetime, resulting in only a marginal performance overhead. This security guarantee of no SAEs in $10^{30}$ years is even stronger than Mirage's guarantee of $10^{17}$ years. 
On average, Avatar-R uses 121 ways per skew and 7 invalid ways per skew for security. In a 16 MB Avatar LLC, this results in a total of 242K (242$\times$1024) valid cache entries and 14K (14$\times$1024) invalid cache entries, leading to a minimal capacity loss of $\approx \! 5\%$.

For randomization, we use the Speck Cipher~\cite{speck}, which is a 32-bit, 22-round lightweight block cipher using 64-bit keys. We use its sister algorithm, Simon~\cite{speck}, for a hardware implementation of Avatar. The cipher, when turned ON, adds an access latency of three cycles for every LLC lookup. To determine the additional latency for accessing a 256-way skewed tag store in Avatar-R compared to a traditional set-associative cache, we used P-CACTI~\cite{PCACTI} with 7nm FinFET technology and CACTI 6.0~\cite{cacti09} with 22nm CMOS technology to model the highly associative design. Both tools show an increase of approximately 0.1ns in access latency, which remains within the 0.25ns cycle time. Even with high associativity, the lookup latency remains essentially unchanged since the tag ways are accessed in parallel. However, this comes at the cost of higher energy consumption, as shown in Table \ref{tab:performance-ass}. In total, Avatar-R requires four additional cycles per lookup when the ciphers are being used.

\subsubsection{Avatar-P}
To mitigate any kind of covert channel, including occupancy-based attacks, one can choose to operate in Avatar-P, with static way partitioning. This prevents the attacker from gaining any knowledge of a secure domain other than its own. We opt for way-based partitioning instead of set-based partitioning as the latter requires OS support. In this mode, the cache skips the cipher, thus eliminating extra 4 cycles of latency and deploying a smart replacement policy restricted to a domain. Avatar-P operates with high associativity, with 256 ways, making it more scalable for servers to deploy. 
This mode can allow the system to partition the cache among a large number of domains (256 domains compared to the 16 domains with the state-of-the-art 16-way way-based partitioned LLC), but this comes at the cost of reduced LLC space per domain.

\subsubsection{Common details for both secure modes}
In Avatar-R and Avatar-P, a 16 MB LLC employs 1024 sets and 256 ways (128 ways per skew, in case of Avatar-R) for an 8-core system, requiring 30 tag bits from the 46-bit line address. Each tag entry includes three additional MOESI coherence protocol coherence bits and three RRIP (unused in Avatar-R) bits. The secure domain ID (SDID) tracks the domain responsible for bringing in a cache line, enabling the duplication of shared cache lines to enhance security against shared-memory attacks. Avatar-R and Avatar-P use four SDID bits to accommodate up to 16 domains, and the number of secure domains supported can be increased by adding more SDID bits to each tag entry. For example, increasing the number of domains to 256 (same as Mirage~\cite{mirage}) will result in an additional storage overhead of 0.8\%. The necessity for SDIDs in Avatar-R and Avatar-P is discussed in Section \ref{sec:shared-memory}. Note that the RRIP bits used for replacement are isolated per domain.
We use 40 bits for each tag, resulting in a tag store size of 1.25 MB. The data store contains 256K (256$\times$1024) entries, each storing 512 bits of data (64 B cache lines). Due to the one-to-one mapping with the tag store, there is no need for a reverse pointer. Thus, 512 bits are stored for each data store entry, yielding a total data store size of 16 MB.


\subsection{On-demand mode switching} 
\label{sec:switch}

The mode switching in Avatar is performed by assigning two MSR bits, with ring zero and exception level-1 privileges for x86 and ARM, respectively, so that only the OS, privileged user, or security monitor~\cite{mi6, sanctum, Cyclone, CIPS} can toggle them. Cyclone~\cite{Cyclone} uses cyclic interference to detect attacks, whereas CIPS~\cite{CIPS} uses a threshold-based detection mechanism. Both these techniques are capable of detecting conflict-based and various other kinds of LLC attacks. Such an attack detector can be used in conjunction with Avatar to dynamically decide which mode to operate the LLC in.

\begin{figure}[htb]
    \centering
    \includegraphics[width=1.0\columnwidth]{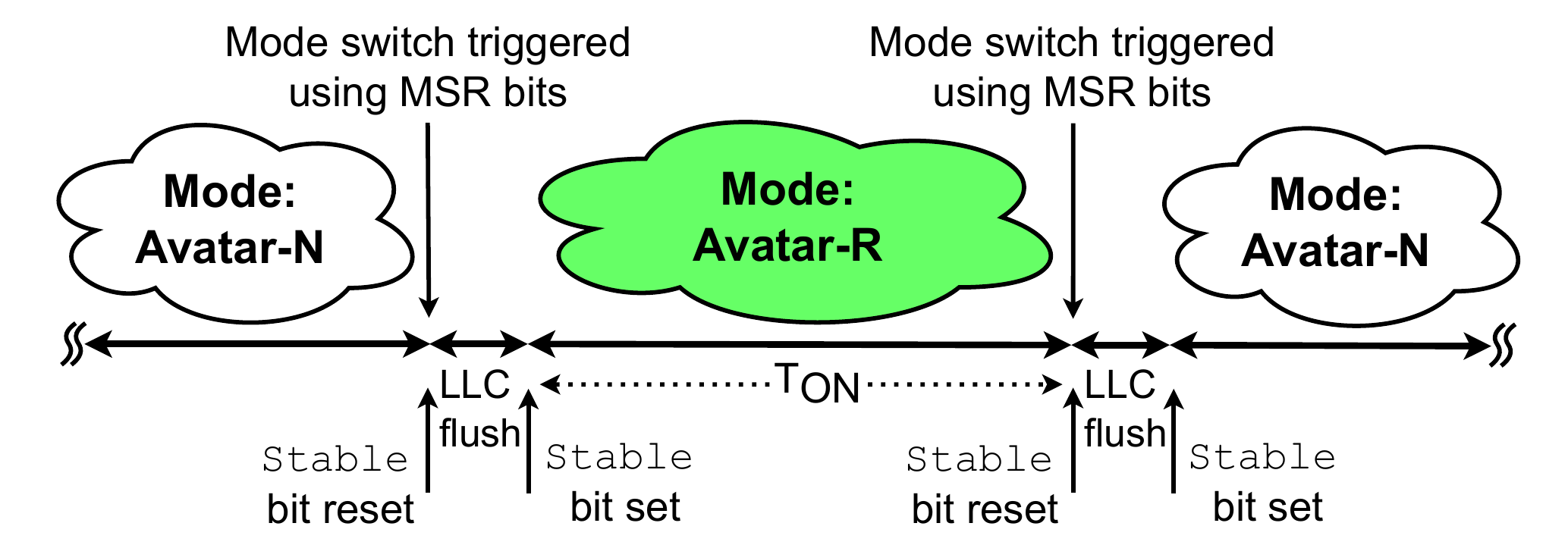}
    \caption{An overview of the steps involved in switching modes in Avatar. Note that this is just one possible example and the system can switch from any mode to any other mode.
    }
    \label{fig:morphing}
\end{figure}

The LLC can be operated in Avatar-N, Avatar-R or Avatar-P mode. The steps for the switching process are shown in Figure \ref{fig:morphing}. 
Note that while the cache is switching modes, LLC accesses are not allowed; otherwise, it can lead to new channels due to mode switching. As the LLC is sliced, for each slice, the cache controller invalidates all LLC lines on a mode switch concurrently. Once the invalidation is complete, the cache controller updates the \texttt{stable} bit, allowing the user to access the LLC. During the mode switch, the \texttt{stable} bit is reset, which prevents any LLC accesses. The user, OS, or security monitor needs to \textit{poll} the \texttt{stable} bit at a fixed interval based on the minimum time required for switching modes. A detailed circuit-level overview of the hardware modifications required to support Avatar’s morphability is provided in Appendix \ref{appendix:A}. It shows the specific LLC components activated in each operating mode (unused hardware can be power-gated to save power).

This runtime mode switching may be more practical in cloud environments. Other dynamic switching methods in Avatar are also viable, with implementation decisions left to the manufacturer based on specific requirements and constraints. In a static switching setup, where the user selects the mode at boot time via BIOS settings, end-user access to this register may also be provided.


\vspace{0.05 in}
\noindent \textbf{Support for Trusted Execution Environments (TEEs).} Switching to Avatar-R or Avatar-P can be done in conjunction with Intel TDX~\cite{tdx} or ARM TrustZone~\cite{arm_trustzone}. In this scenario, when an application enters a secure enclave, Avatar transitions to one of these secure modes and invalidates the LLC.

\begin{figure}[htb]
    \centering
    \vspace{-0.05 in}    
    \includegraphics[width=0.98\columnwidth]{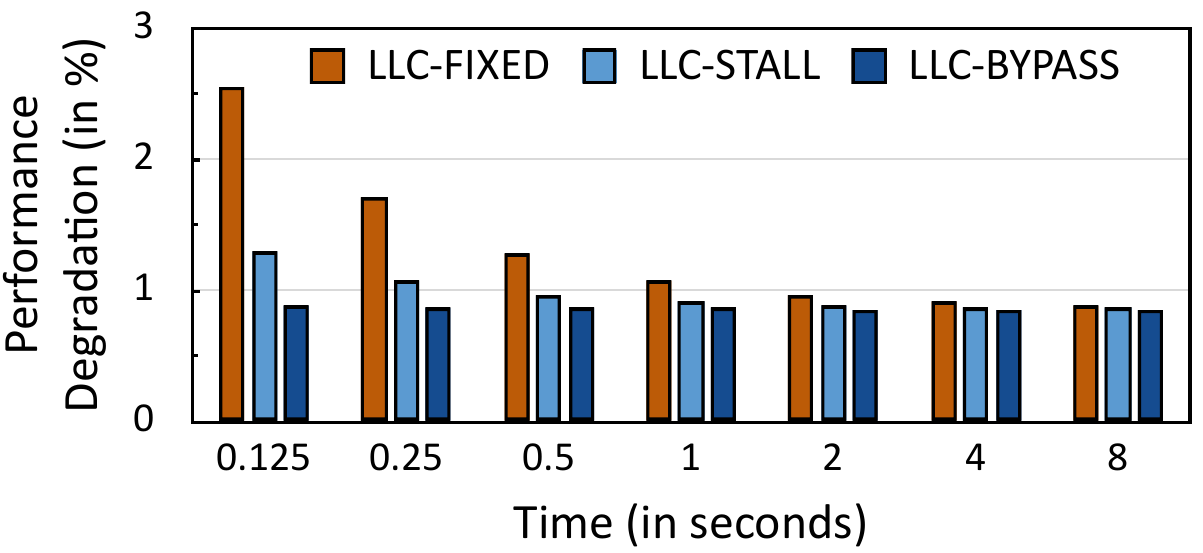}
    \vspace{-0.05in}
    \caption{Performance degradation with varying ${T_{ON}}$ between two consecutive mode switches (for SPEC CPU2017).}
    \label{fig:switching}
\vspace{-0.00 in}
\end{figure}

\vspace{0.05 in}
\noindent \textbf{DoS attacks.}
We mandate a minimum operation time ${T_{ON}}$ of one-second for any mode in Avatar. This is to prevent denial-of-service attacks in which the attacker continuously switches between modes and keeps the LLC occupied. ${T_{ON}}$ can be made shorter for better responsiveness in TEE and cloud environments. However, this comes with additional performance degradation, assuming there is a switch after every ${T_{ON}}$ seconds. We explore this in detail in Figure~\ref{fig:switching}.

\vspace{0.05 in}

\noindent \textbf{Switching overhead.} The overhead for switching depends on how often we switch between modes in a fixed time interval while running our application of interest. We analyze the impact on the system's performance based on the operation time of a secure mode. During a switch, the LLC must be flushed so that no information about the secure execution is leaked to an adversary. Figure \ref{fig:morphing} shows a typical switching cycle where the system changes mode twice with an operation time of ${T_{ON}}$.

We quantify the switching overhead for a range of operation times ${T_{ON}}$, as shown in Figure \ref{fig:switching}. 
We analyze three approaches for performing LLC flushing during mode switching. In the LLC-STALL approach, all LLC accesses are halted during the transition. This causes the L1 and L2 caches to stall on misses, resulting in temporary performance degradation. Alternatively, the LLC-BYPASS approach allows continued LLC access during the switch. As flushing progresses, the LLC hit rate gradually declines, eventually reaching a 100\% miss rate—effectively mimicking a system with no LLC. In this case, we approximate it to a system with a completely bypassed LLC to get an upper bound of degradation. Both these approaches depend on the number of dirty lines in the LLC that need to be written back during LLC flush. To avoid leaking this information about the secret application running, we also show the LLC-FIXED approach, wherein the LLC writes back all cache lines to the main memory. This makes sure that the flush latency is independent of the number of write-backs in the application at the time of switching modes. Figure~\ref{fig:switching} illustrates the performance overhead of all approaches as the operation time ($T_{ON}$) varies from one-eighth of a second to eight seconds. 

\begin{figure}[htb]
    \centering
    \includegraphics[width=0.98\linewidth]{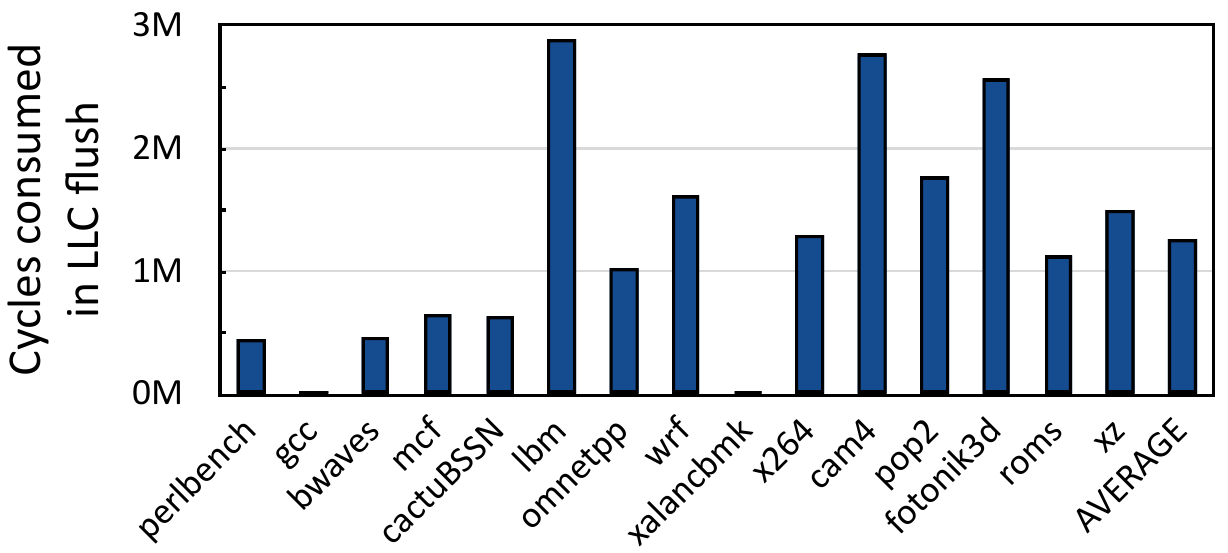}
    \caption{CPU cycles consumed in LLC flush during a mode switch (for SPEC CPU2017). The average cycle count is around 1.25 million cycles.}
    \vspace{-0.15 in}
    \label{fig:flush-cycles}
\end{figure}

\begin{figure}[htb]
    \centering
    \includegraphics[width=0.98\linewidth]{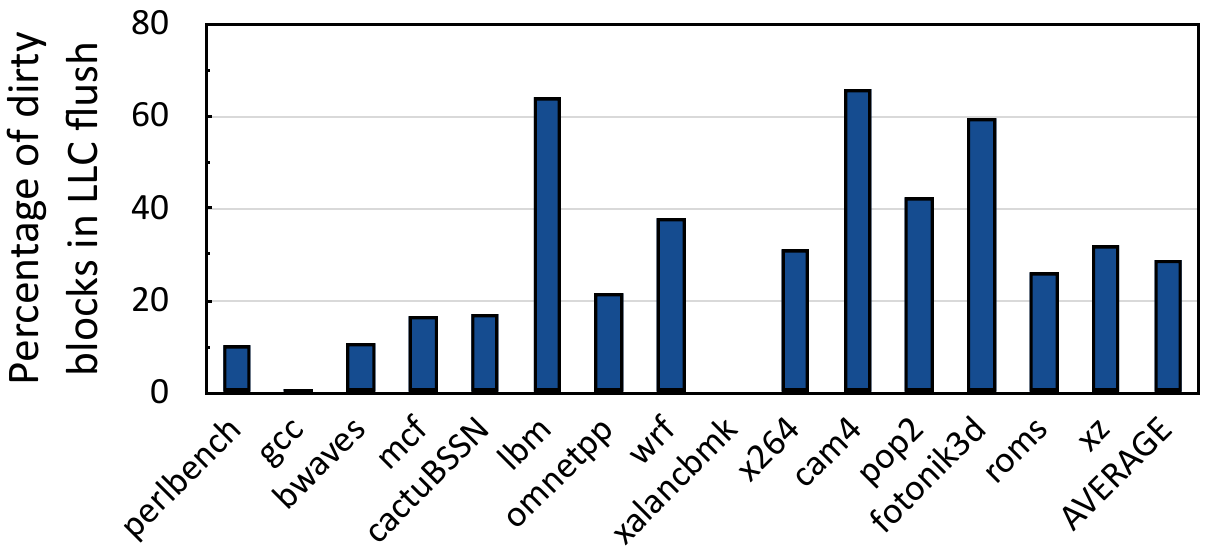}
    \vspace{-0.05 in}
    \caption{Percentage of blocks that cause writebacks during an LLC flush (for SPEC CPU2017). On average, around 29\% of blocks are dirty.}
    \label{fig:writeback_percentage}
    \vspace{-0.00 in}
\end{figure}

The LLC-STALL approach is recommended for scenarios where a potential flush latency-based attack is not a concern, as it is simpler to manage in hardware, and the performance difference between these methods vanishes beyond $T_{ON}$ of one second. The LLC flushing latency is primarily constrained by the number of read ports per slice and the DRAM writeback bandwidth. As shown in Figure~\ref{fig:flush-cycles}, the average flush latency is approximately 1.25 million cycles, significantly smaller than the one-second (four billion cycle) cooldown period. Furthermore, as illustrated in Figure~\ref{fig:writeback_percentage}, the flush time is dictated by the proportion of lines that must be written to DRAM. The more robust LLC-FIXED approach ensures a flush latency of approximately 4.33 million cycles by forcing all LLC lines (clean or dirty) to be written back to main memory. This worst-case measure hides application-dependent latencies and mitigates potential side-channels.





\subsection{Avatar and private caches}
In modern processors, private L1 and L2 caches are often shared by 2-way simultaneous multi-threading (SMT) cores. Partitioning these caches between threads is a common and straightforward design choice, as shown in prior work~\cite{nomo}. The 4-cycle access latency of Avatar-R, combined with high associativity and global random replacement at L1, can introduce significant performance overhead. Thus, partitioning these private caches is advisable. Similarly, the cache coherence directory can also be partitioned~\cite{secdir}. 

\section{Security analysis of Avatar-R}
\label{sec:security}
Conflict-based attacks rely on creating eviction sets to leak secret information, typically using algorithms like Prime, Prune and Probe~\cite{systematicAnalysis}, Conflict Testing~\cite{conflictTesting}, and Prune+Plumtree~\cite{plumtree}. These methods fundamentally depend on successful SAEs. While secure cache designs often evaluate their resilience against these known algorithms, undiscovered or faster algorithms could potentially create eviction sets with even fewer SAEs, threatening security. To address this, we show that Avatar-R makes even a single SAE exceedingly improbable over the system’s lifetime, thereby providing robust security (Appendix~\ref{appendix:B}).

\subsection{Requirements on randomizing function}
The randomizing function used to determine the set mapping in Avatar-R is crucial for achieving a balanced distribution of invalid entries across sets, thus eliminating SAEs. For our analysis, we assume perfectly random set-mapping functions, ensuring uniform cache line distribution. However, recent shortcut attacks~\cite{probabilistic2} have demonstrated the potential to exploit vulnerabilities in these algorithms to create deterministic collisions. To counter such attacks, we employ a cryptographic function computed in hardware with a secret key that remains unknown to attackers. Additionally, each skew utilizes an independent mapping function with a unique secure key, ensuring mutual independence of the mapping functions. A correct implementation of this hardware cipher is also crucial to the security properties of a secure randomized cache, as highlighted in~\cite{kgp-response}. We use the bucket and balls model to estimate the probability of an SAE. Avatar-R with 128 ways per skew has a frequency of one SAE in $5 \cdot 10^{46}$ line installs or once in around $10^{30}$ years, effectively providing even stronger security than Mirage against conflict-based attacks. Please refer to Appendix~\ref{appendix:B} for detailed calculations.  

{
\setlength{\tabcolsep}{3pt}
\begin{table}[htb]
  \caption{LLC line installs per SAE as the associativity of Avatar-R varies from 128 ways to 512 ways. Note that the LLC size is kept constant at 16 MB, and seven ways per skew are invalid.}
    \centering
    \renewcommand{\arraystretch}{1.5}
   \resizebox{!}{0.6cm}{
    \begin{tabular}{|c||c|c|c|}
    \hline
        \textbf{Associativity} & \textbf{128-ways} & \textbf{256-ways} & \textbf{512-ways} \\ \hline \hline
        \textbf{Installs} & $10^{50}$ ($10^{34}$ yrs) & $10^{46}$ ($10^{30}$ yrs) & $10^{44}$ ($10^{28}$ yrs) \\
    \hline
    \end{tabular}
    }
    \label{tab:security-ass}
\end{table}
}

\subsection{Sensitivity to associativity}
We now vary the associativity of Avatar-R, keeping the LLC size at 16 MB. The associativity varies from 128 to 512 ways, with the default configuration having 256 total ways (128 ways per skew). Seven ways per skew are kept invalid for all these Avatar-R configurations. Table \ref{tab:security-ass} shows the rate of SAE for these configurations. The 128-way configuration is the most secure (one SAE in $10^{34}$ years), and security reduces as the LLC associativity increases. However, even for the 512-way configuration, the rate of SAE is once in $10^{28}$ years.

\subsection{Effect of no decoupling on security}
\label{sec:decoupling}

Mirage uses a decoupled tag and data store design with pointer-based indirection between the tag and data entries. Mirage's use of a decoupled tag-data design arises from additional invalid tag entries, which creates a mismatch between the number of tags and data entries, necessitating tag-data indirection. However, our security simulations demonstrate that removing this decoupling while keeping the extra invalid tags as unused entries has no impact on the security of Avatar-R or Mirage, assuming the tag and data stores contain an equal number of entries. A recent study~\cite{sok} also showed that decoupling the tag and data store to accommodate invalid tags has no security impact.

\subsection{Key management}
The key used in Avatar-R is established during system boot. The keys are stored in hardware and are not visible to any software, including the OS. Avatar-R does not require a continuous refresh of the keys, as no information about the set mapping can be leaked without SAEs. SAEs are easily detected by monitoring invalid entries in each skew as part of the load-aware insertion policy. If both skews are full, an entry is randomly evicted from one skew, causing an SAE. Upon detecting an SAE, the cache can be flushed and re-keyed. Our analysis shows that the probability of an SAE in Avatar-R is extremely low, with re-keying likely only if an attacker brute-forces the key (one in $2^{64}$ probability). 

\subsection{Need for secure domain IDs}
\label{sec:shared-memory}
In situations where the attacker and victim do share LLC lines, various attacks like Flush+Reload~\cite{flush+reload}, Flush+Flush~\cite{flush+flush}, Flush+Prefetch \cite{flush+prefetch}, and Evict+Reload~\cite{evict+reload}, could potentially leak victim data. Avatar-R includes a 4-bit SDID for each tag entry, allowing up to 16 domains. This design ensures that LLC fills from a secure domain don't impact those from another, thus providing security against shared-memory attacks. Avatar maintains the existing protocol for cache coherence but includes the SDID in all coherence packets. This is only applicable to a shared read-only memory, as a shared writeable memory is incompatible with cache coherence protocols~\cite{DAWG, scattercache}. The SDID length can be adjusted to support different numbers of domains as needed. Avatar-R also mitigates the Reload+Refresh~\cite{reload+refresh} attack as it guarantees global evictions with random replacement.

\begin{figure*}[htp]
    \centering
    \includegraphics[width=0.98\textwidth]{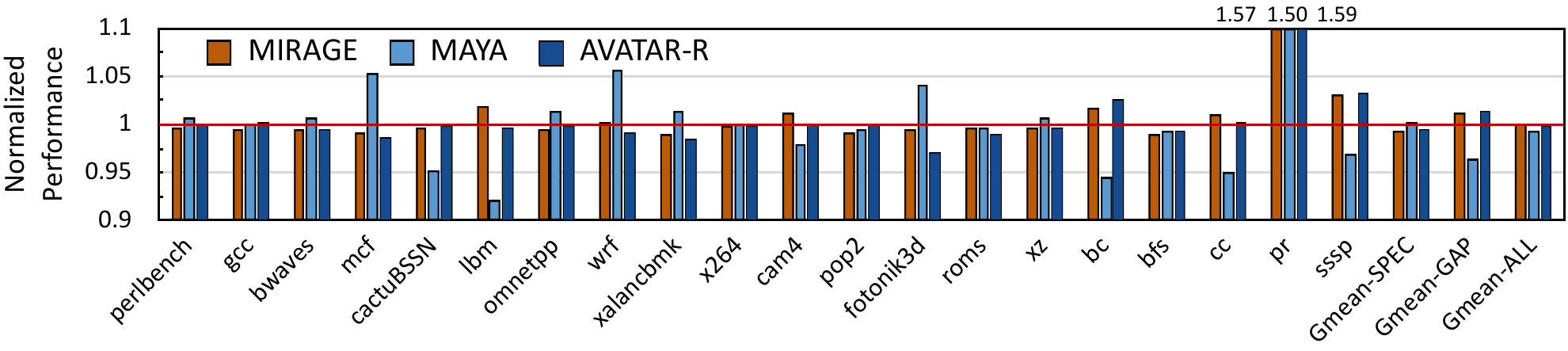}
    \vspace{-0.1 in}
    \caption{Performance of Avatar-R for 8-core homogeneous mixes (from SPEC CPU2017~\cite{spec} and GAP~\cite{gap} suites), normalized to the non-secure baseline. The geomeans exclude \texttt{pr}. Avatar-R and Maya have a marginal performance overhead of 0.2\% and 0.6\%, respectively.}
    \label{fig:homoperf}
   \vspace{-0.05 in}
\end{figure*}

\begin{figure*}[htp]
    \centering
    \includegraphics[width=0.98\textwidth]{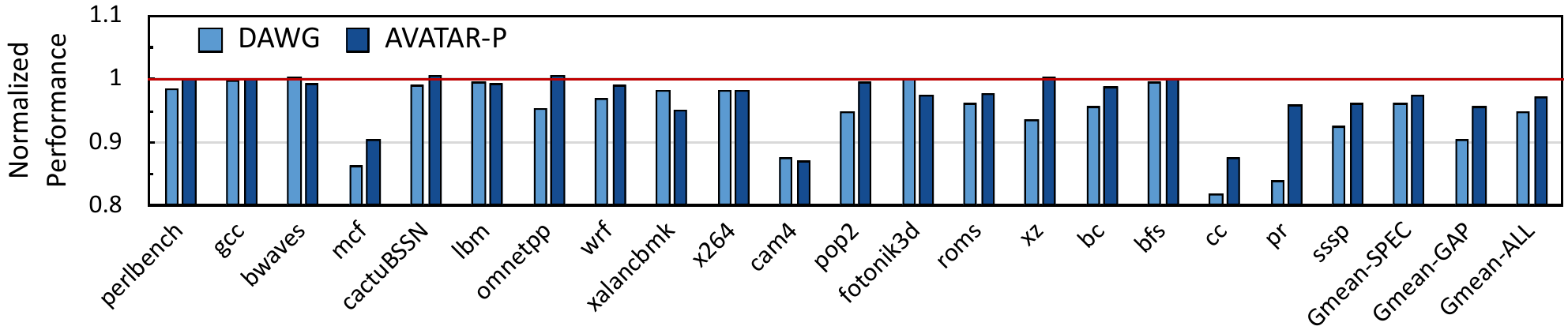}   
    \vspace{-0.1 in}
    \caption{Performance of Avatar-P (256-way static way-based partitioning) for 8-core homogeneous mixes (from SPEC CPU2017~\cite{spec} and GAP~\cite{gap} suites), normalized to the non-secure baseline. Avatar-P outperforms the 16-way 16 MB DAWG-based partitioning, with an overhead of less than 3\% compared to the non-secure baseline.}
    \label{fig:partitioning-detailed}
    \vspace{-0.05 in}
\end{figure*}

\section{Performance evaluation}
\label{sec:evaluation}

\vspace{-0.3em}
\subsection{Methodology}
\label{sec:methodology}
We evaluate different LLC designs using the ChampSim~\cite{champsim} microarchitecture simulator. Building on the artifact provided by Maya~\cite{maya}, we modify it to implement the Avatar LLC. Our baseline is a non-secure 8-core system with a 16 MB, 16-way set-associative LLC, and 64-byte cache lines (Table \ref{tab:baseline}).
We test 15 homogeneous workloads (using 42 sim-points) from SPEC CPU2017~\cite{spec} and five homogeneous workloads (using 20 sim-points) from GAP~\cite{gap}, selecting benchmarks with more than one LLC miss per kilo instruction (MPKI) for the baseline configuration. Benchmarks are chosen based on their LLC MPKI for a single-core 2 MB LLC. Additionally, we evaluate 21 heterogeneous mixes with 8 randomly selected benchmarks from the SPEC CPU2017 and GAP suites running on the 8-core system.
Simulations run 1.6B instructions across eight cores (200M per core) in the region of interest, following a 50M instruction warmup per core.
To assess performance, we use the weighted speedup~\cite{weight} metric for an 8-core system. First, we compare Avatar-R against the non-secure baseline (equivalent performance to Avatar-N). Then, we evaluate the performance gains of Avatar-P over state-of-the-art way-based partitioned LLC (DAWG~\cite{DAWG}).
{
\setlength{\tabcolsep}{4pt}
\begin{table}[t]
  \caption{Simulation parameters of the baseline system.}
  \centering
  \renewcommand{\arraystretch}{1.5}
  \resizebox{!}{2.5cm}{
  \begin{tabular}{|c||l|} \hline
  \multicolumn{1}{|c||}{\multirow{2}{*}{\begin{tabular}[c]{@{}c@{}}\textbf{Core}\end{tabular}}} & \multicolumn{1}{l|}{\multirow{2}{*}{\begin{tabular}[l]{@{}l@{}}8 cores, Out-of-order, bimodal~\cite{bimodal}, 4~GHz with 6-issue\\[-0.4em]width, 4-retire width, 512-entry ROB\end{tabular}}} \\
  \multicolumn{1}{|c||}{} & \multicolumn{1}{c|}{} \\ \hline
  \multicolumn{1}{|c||}{\multirow{2}{*}{\begin{tabular}[c]{@{}c@{}}\textbf{TLBs}\end{tabular}}} & \multicolumn{1}{l|}{\multirow{2}{*}{\begin{tabular}[l]{@{}l@{}}L1 ITLB/DTLB: 64 entries, 4-way, 1 cycle, STLB: 2048\\[-0.4em]entries, 16-way, 8 cycles\end{tabular}}} \\
  \multicolumn{1}{|c||}{} & \multicolumn{1}{c|}{} \\ \hline
  \textbf{L1I} & 32~KB, 8-way, 1 cycle, LRU \\ \hline
  \textbf{L1D} & 48~KB, 12-way, 5 cycles, LRU, IPCP prefetcher~\cite{ipcp} \\ \hline
  \textbf{L2} & 512~KB 8-way, 10 cycles, LRU, non-inclusive, IPCP pref.\\ \hline
  \textbf{LLC} & 2~MB/core, 16-way, 24 cycles, Hawkeye~\cite{Jain2017HawkeyeL}, non-inclusive\\ \hline
  \textbf{MSHRs} & 8/16/32 at L1I/L1D/L2, 64/core at the LLC \\ \hline
  \multicolumn{1}{|c||}{\multirow{2}{*}{\begin{tabular}[c]{@{}c@{}}\textbf{DRAM}\\[-0.4em]\textbf{controller}\end{tabular}}} & \multicolumn{1}{l|}{\multirow{2}{*}{\begin{tabular}[l]{@{}l@{}}DDR4-3200, two channels/8-cores 4~KB row-buffer per\\[-0.4em]bank, open page, burst length 16, t\textsubscript{RP, RCD, CAS}: 12.5~ns\end{tabular}}} \\
   & \\ \hline
  \end{tabular}
  }
  \label{tab:baseline}
  \vspace{-1.2em}
\end{table}
}




\subsection{Avatar-R performance}

\noindent \textbf{Homogeneous mixes.} Figure \ref{fig:homoperf} illustrates the performance of the Avatar-R normalized to the non-secure baseline for various homogeneous SPEC CPU2017 and GAP workloads. On average, for the SPEC CPU2017 benchmarks, the Avatar-R experiences a marginal performance loss of 0.6\% compared to the non-secure baseline. However, Avatar-R incurs large performance slowdowns for specific benchmarks, such as 
\texttt{fotonik3d}. 
\texttt{fotonik3d} is a prefetcher-sensitive application, which means the prefetcher significantly aids in reducing MPKI for \texttt{fotonik3d}, but the global random replacement policy of Avatar-R degrades the prefetcher's performance. However, such a policy is critical to the security of the Avatar cache, and cannot be replaced by a smarter replacement policy.

For the \texttt{pr} workload, Avatar-R, Maya, and Mirage outperform the non-secure baseline by 59\%, 50\%, and 57\%, respectively. This is due to a weak baseline, where the IPCP prefetcher negatively affects Hawkeye's performance, making it worse than using no prefetching with the LRU policy. For the remaining benchmarks, Avatar-R performs quite similarly to Mirage. Since \texttt{pr} is an outlier, we exclude it from geomean performance calculations. Without \texttt{pr}, Avatar-R performs 1.3\% better than the non-secure baseline for GAP workloads. Maya has a 3.6\% performance slowdown on GAP workloads (excluding \texttt{pr}) due to its smaller data store. The geomean across SPEC CPU2017 and GAP workloads (excluding \texttt{pr}) shows that Avatar-R incurs a minimal 0.25\% overhead. 

\vspace{0.05 in}
\noindent \textbf{Heterogeneous mixes.} For heterogeneous workloads, Avatar-R incurs a 1\% average performance degradation compared to the non-secure baseline. This is due to the 4-cycle additional access latency and the global random replacement policy, same as Mirage. However, it performs slightly worse than Mirage due to its smaller usable data store (95\% of total capacity).

\vspace{0.05 in}
\noindent \textbf{Sensitivity to cipher latency.}
\label{sec:cipher}
For Avatar-R, we use a 32-bit, 22-round lightweight Simon cipher~\cite{speck} with 64-bit keys and a 3-cycle latency. 
We evaluate Avatar-R's performance for SPEC CPU2017 workloads across cipher latencies ranging from one to five cycles. We observe that with a 1-cycle cipher the slowdown for Avatar-R is 0.4\%, and 0.8\% with a 5-cycle cipher.

\vspace{0.05 in}
\noindent \textbf{Impact of global random eviction on performance.}
Global random eviction is key to ensuring strong security in Avatar-R but can affect performance. However, state-of-the-art replacement policies like Hawkeye are not effective for GAP workloads~\cite{p-opt}, and the usage of a global random eviction policy does not affect LLC performance. It is important to note that this does not undermine the effectiveness of smart replacement policies. Given that evictions are performed globally across the entire cache, the expanded pool of candidates diminishes the likelihood of selecting the worst candidate for eviction, in contrast to a random replacement policy limited to a set where the likelihood of selecting the worst possible candidate is much higher.
Avatar-R uses skews that reduce conflict misses and improve cache performance~\cite{seznec93}, amortizing the negative effect of the global random eviction policy.



\subsection{Avatar-P performance}
\label{sec:perf-partitioning}
Traditional partitioning techniques face high performance overheads and scalability challenges. We show that Avatar-P significantly reduces these overheads, making it a practical defense against both occupancy- and eviction-based attacks.

\vspace{0.05 in}
\noindent \textbf{Homogeneous mixes.}
Figure \ref{fig:partitioning-detailed} demonstrates that Avatar-P achieves notable performance improvements over state-of-the-art 16-way DAWG-based partitioned LLC when used with way-based partitioning. We note that memory-intensive workloads such as \texttt{mcf} and \texttt{cam4}, and graph workloads such as \texttt{cc} and \texttt{pr} benefit greatly from higher available LLC ways/set, reducing conflict misses. On average, Avatar-P has a 3\% performance overhead compared to the non-secure baseline. 

\vspace{0.05 in}

\noindent \textbf{Heterogeneous mixes.} We also evaluate Avatar-P on heterogeneous mixes comprising SPEC CPU2017 and GAP workloads. On average, Avatar-P incurs a 4\% performance degradation compared to the non-secure baseline. This overhead is primarily due to the static partitioning used in Avatar-P, which can lead to increased conflict misses, particularly for memory-intensive workloads.

\vspace{0.05 in}

\noindent \textbf{Support for large number of domains.} 
The default Avatar-P design employs way-based partitioning to split the cache across security domains. By default, the system supports up to 16 domains, but this can be expanded by increasing the number of tag bits. For instance, supporting 64 security domains would require 6 SDID bits per tag entry, incurring only a 0.3\% additional storage overhead. However, way-based partitioning does not scale efficiently to such a large number of domains, even with high associativity. Therefore, we recommend switching to a more sophisticated set-based partitioning algorithm in Avatar-P to minimize performance degradation.

Avatar-P delivers significant performance gains even with set-based partitioning compared to state-of-the-art 16-way set-based partitioned LLCs (e.g., BCE and page coloring). For example, Avatar-P with BCE outperforms a 16-way BCE-partitioned LLC by approximately 12\%, and similar improvements are observed when using page coloring.

\begin{figure}[htb]
    \centering
    \includegraphics[width=\linewidth]{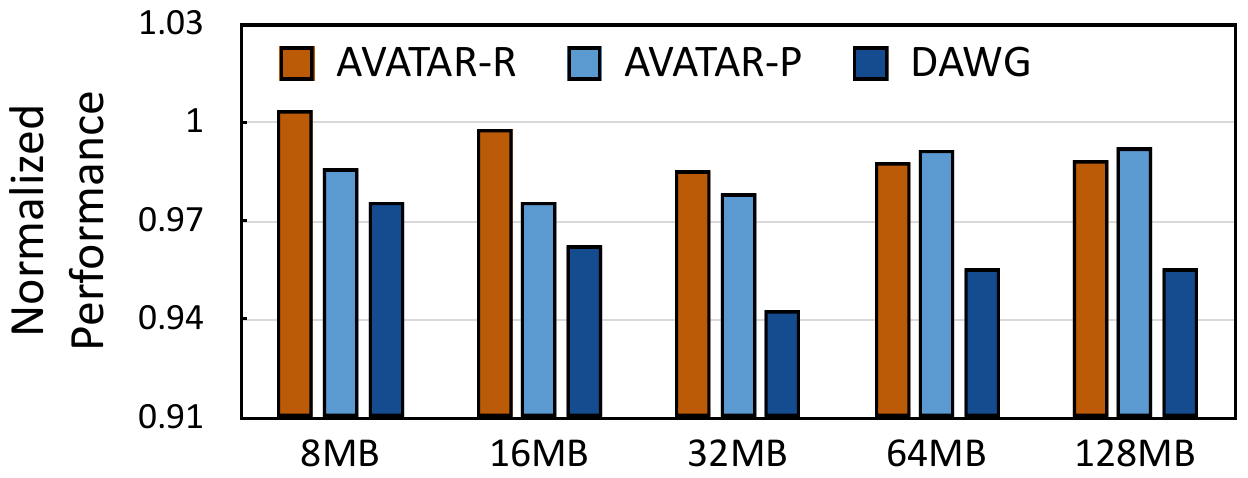}   
    \caption{Performance of Avatar-R, Avatar-P, and 16-way 16 MB 8-core DAWG-based partitioned LLC with LLC size varying from 8MB to 128MB. 
    }
    \label{fig:cache-size}
    \vspace{-0.05 in}
\end{figure}

\subsection{Sensitivity to LLC size}
We evaluated Avatar-R and Avatar-P using SPEC CPU2017 workloads across varying LLC sizes, ranging from 8 MB to 128 MB (corresponding to 1–16 MB per core), extending beyond the default 16 MB (2 MB per core) Avatar LLC configuration. 
Figure~\ref{fig:cache-size} shows the resulting performance trends.
At 8 MB, Avatar-R slightly outperforms its non-secure baseline by approximately 0.3\%. As LLC size increases beyond 32 MB, the performance advantage diminishes, stabilizing at about a 1\% overhead due to the higher LLC hit rate and reduced opportunity for further gains. This demonstrates that Avatar-R remains practical even for systems with large LLCs.
Avatar-P consistently outperforms the state-of-the-art DAWG-based way-partitioned LLC across all cache sizes, with the performance gap increasing at larger sizes. Notably, Avatar-P begins to surpass Avatar-R at 64 MB (8 MB per core). This suggests that Avatar-P may be preferable for securing workloads on systems with very large LLC/core, which is typically not the case with commercial processors, as 2-4 MB/core is the current trend.



\section{Storage and power overheads}
\label{sec:Overheads}
\noindent \textbf{Storage.}
As detailed in Section \ref{sec:implementation}, Avatar requires 40 tag bits for its morphable implementation, compared to 32 tag bits in the non-secure baseline. This increases Avatar’s total tag store size to 1280 KB, up from 1024 KB in the baseline. Avatar and the baseline have a data store of 512 bits per entry, resulting in a total storage requirement of 16,384 KB. Consequently, Avatar’s overall storage requirement is 17,664 KB, while the baseline requires 17,408 KB, leading to a 1.5\% storage overhead for Avatar.

{
\begin{table}[htp]
  \caption{Energy per access (E/A) and power consumption for the non-secure baseline, Mirage, Avatar-N, and Avatar-R/Avatar-P.}
    \centering
    \renewcommand{\arraystretch}{1.3}
    \resizebox{!}{3.0cm}{
    \begin{tabular}{|c||c|c|c|c|}
    \hline
        \textbf{Design} & \begin{tabular}[c]{@{}c@{}}\textbf{Baseline}\end{tabular} & \begin{tabular}[c]{@{}c@{}}\textbf{Mirage}\end{tabular} & \begin{tabular}[c]{@{}c@{}}\textbf{Avatar-N}\end{tabular} & \begin{tabular}[c]{@{}c@{}}\textbf{Avatar-R} \\[-0.3em] \textbf{/ Avatar-P}\end{tabular} \\ \hline \hline

        \multicolumn{1}{|c||}{\multirow{2}{*}{\begin{tabular}[c]{@{}c@{}}Tag Read\\[-0.3em] E/A (nJ)\end{tabular}}} & \multicolumn{1}{c|}{\multirow{2}{*}{\begin{tabular}[c]{@{}c@{}}0.007\end{tabular}}} & \multicolumn{1}{c|}{\multirow{2}{*}{\begin{tabular}[c]{@{}c@{}}0.011\end{tabular}}} & \multicolumn{1}{c|}{\multirow{2}{*}{\begin{tabular}[c]{@{}c@{}}0.007\end{tabular}}} & \multicolumn{1}{c|}{\multirow{2}{*}{\begin{tabular}[c]{@{}c@{}}0.026\end{tabular}}} \\
        \multicolumn{1}{|c||}{} & \multicolumn{1}{c|}{} & \multicolumn{1}{c|}{} & \multicolumn{1}{c|}{} & \multicolumn{1}{c|}{} \\ \hline

        \multicolumn{1}{|c||}{\multirow{2}{*}{\begin{tabular}[c]{@{}c@{}}Data Read\\[-0.3em] E/A (nJ)\end{tabular}}} & \multicolumn{1}{c|}{\multirow{2}{*}{\begin{tabular}[c]{@{}c@{}}0.164\end{tabular}}} & \multicolumn{1}{c|}{\multirow{2}{*}{\begin{tabular}[c]{@{}c@{}}0.170\end{tabular}}} & \multicolumn{1}{c|}{\multirow{2}{*}{\begin{tabular}[c]{@{}c@{}}0.164\end{tabular}}} & \multicolumn{1}{c|}{\multirow{2}{*}{\begin{tabular}[c]{@{}c@{}}0.164\end{tabular}}} \\
        \multicolumn{1}{|c||}{} & \multicolumn{1}{c|}{} & \multicolumn{1}{c|}{} & \multicolumn{1}{c|}{} & \multicolumn{1}{c|}{} \\ \hline

        \multicolumn{1}{|c||}{\multirow{2}{*}{\begin{tabular}[c]{@{}c@{}}Tag Write\\[-0.3em] E/A (nJ)\end{tabular}}} & \multicolumn{1}{c|}{\multirow{2}{*}{\begin{tabular}[c]{@{}c@{}}0.018\end{tabular}}} & \multicolumn{1}{c|}{\multirow{2}{*}{\begin{tabular}[c]{@{}c@{}}0.048\end{tabular}}} & \multicolumn{1}{c|}{\multirow{2}{*}{\begin{tabular}[c]{@{}c@{}}0.018\end{tabular}}} & \multicolumn{1}{c|}{\multirow{2}{*}{\begin{tabular}[c]{@{}c@{}}0.101\end{tabular}}} \\
        \multicolumn{1}{|c||}{} & \multicolumn{1}{c|}{} & \multicolumn{1}{c|}{} & \multicolumn{1}{c|}{} & \multicolumn{1}{c|}{} \\ \hline

        \multicolumn{1}{|c||}{\multirow{2}{*}{\begin{tabular}[c]{@{}c@{}}Data Write\\[-0.3em] E/A (nJ)\end{tabular}}} & \multicolumn{1}{c|}{\multirow{2}{*}{\begin{tabular}[c]{@{}c@{}}0.174\end{tabular}}} & \multicolumn{1}{c|}{\multirow{2}{*}{\begin{tabular}[c]{@{}c@{}}0.180\end{tabular}}} & \multicolumn{1}{c|}{\multirow{2}{*}{\begin{tabular}[c]{@{}c@{}}0.174\end{tabular}}} & \multicolumn{1}{c|}{\multirow{2}{*}{\begin{tabular}[c]{@{}c@{}}0.174\end{tabular}}} \\
        \multicolumn{1}{|c||}{} & \multicolumn{1}{c|}{} & \multicolumn{1}{c|}{} & \multicolumn{1}{c|}{} & \multicolumn{1}{c|}{} \\ \hline

        \multicolumn{1}{|c||}{\multirow{2}{*}{\begin{tabular}[c]{@{}c@{}}Tag Static\\[-0.3em] Power (mW)\end{tabular}}} & \multicolumn{1}{c|}{\multirow{2}{*}{\begin{tabular}[c]{@{}c@{}}31\end{tabular}}} & \multicolumn{1}{c|}{\multirow{2}{*}{\begin{tabular}[c]{@{}c@{}}122\end{tabular}}} & \multicolumn{1}{c|}{\multirow{2}{*}{\begin{tabular}[c]{@{}c@{}}45\end{tabular}}} & \multicolumn{1}{c|}{\multirow{2}{*}{\begin{tabular}[c]{@{}c@{}}45\end{tabular}}} \\
        \multicolumn{1}{|c||}{} & \multicolumn{1}{c|}{} & \multicolumn{1}{c|}{} & \multicolumn{1}{c|}{} & \multicolumn{1}{c|}{} \\ \hline

        \multicolumn{1}{|c||}{\multirow{2}{*}{\begin{tabular}[c]{@{}c@{}}Data Static\\[-0.3em] Power (mW)\end{tabular}}} & \multicolumn{1}{c|}{\multirow{2}{*}{\begin{tabular}[c]{@{}c@{}}489\end{tabular}}} & \multicolumn{1}{c|}{\multirow{2}{*}{\begin{tabular}[c]{@{}c@{}}507\end{tabular}}} & 
        \multicolumn{1}{c|}{\multirow{2}{*}{\begin{tabular}[c]{@{}c@{}}489\end{tabular}}} & 
        \multicolumn{1}{c|}{\multirow{2}{*}{\begin{tabular}[c]{@{}c@{}}489\end{tabular}}} \\
        \multicolumn{1}{|c||}{} & \multicolumn{1}{c|}{} & \multicolumn{1}{c|}{} & \multicolumn{1}{c|}{} & \multicolumn{1}{c|}{} \\ \hline
        
    \hline
    \end{tabular}
    }
    \label{tab:power}
\end{table}
}

\noindent \textbf{Power consumption and energy.}
We use 7 nm FinFET technology, simulated using P-CACTI~\cite{PCACTI} in sequential access mode, to estimate the static power and the dynamic access energy. Table \ref{tab:power} summarizes the observed dynamic energy and static power results for an 8-core 16 MB Avatar LLC design in Avatar-N and Avatar-R/Avatar-P modes and compares these results to those of Mirage and the non-secure baseline.
Avatar's static power marginally increases by 2.7\% compared to the non-secure baseline. This increase is due to the additional eight tag bits Avatar uses for each tag entry. In contrast, Mirage incurs a static power overhead of 20.96\%, owing to the larger tag store and same-sized data store compared to the non-secure baseline.

\begin{figure}[htb]
    \centering
    \includegraphics[width=0.48\textwidth]{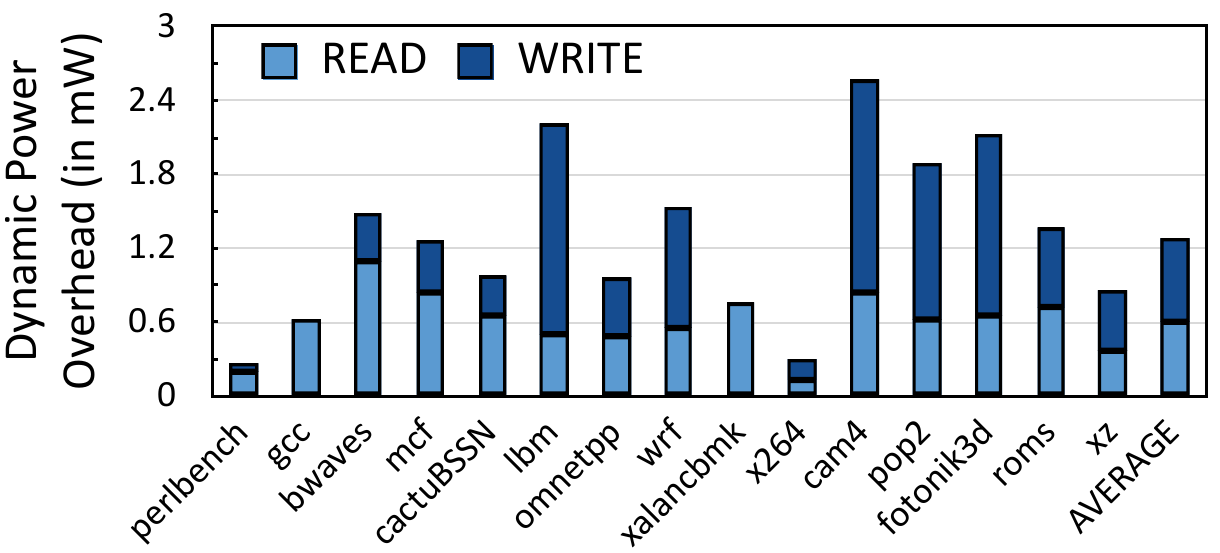}   
    \vspace{-0.05 in}
    \caption{Dynamic read and write power overheads of Avatar-R/Avatar-P as compared to the non-secure baseline. The average dynamic power overhead is 1.28 mW, which is only around 0.2\% of the LLC static power.}
    \label{fig:readwrite}
\end{figure}

We observe that the tag read energy increases by $3.7\times$ in Avatar-R and Avatar-P due to their higher associativity. However, since data read energy remains unchanged and dominates overall energy consumption, this results in only an 11.11\% increase in total dynamic read energy. The tag write energy rises by $5.6\times$, leading to a 43.23\% increase in total dynamic write energy for Avatar-R and Avatar-P compared to Avatar-N and the non-secure baseline. The static power remains the same for all three operating modes because the same hardware is used. This significant increase in dynamic access energy is due to increased associativity (a more complex tag lookup). Figure \ref{fig:readwrite} shows the read and write power overhead of Avatar-R/Avatar-P over the non-secure baseline across various homogeneous SPEC CPU2017 workloads. On average, the read power overhead is around 0.60 mW, and the write power overhead is around 0.68 mW. These are negligible ($\approx \! 0.2\%$) compared to the static LLC power. 


\vspace{0.05 in}
\noindent \textbf{Logic.} The Avatar LLC design requires two modules of the Simon block cipher~\cite{speck}. Recent works~\cite{simon, simon2} have shown efficient hardware implementations of the Simon cipher and evaluated their hardware costs. For 7 nm FinFET technology, the two modules of the Simon cipher have an estimated hardware cost of 6,160 Gate Equivalents (GEs, number of equivalent 2-input NAND gates) and an area requirement of around 61.6 $\mu m^2$. Avatar also requires extra logic for highly associative cache lookups and load-aware skew selection in the randomized secure mode. The additional logic gates that facilitate the morphing ability of Avatar between its operating modes (Figure \ref{fig:oasis-circuit}) amount to around 100,000 GEs or 1,000 $\mu m^2$. The load-aware skew-selection circuit (counting 1s among valid bits of 128 tags from the indexed set in each skew, followed by an 8-bit comparison) requires around 1,660 GEs (16.6 $\mu m^2$). All the extra logic required for the secure operation modes and morphability of Avatar can fit in less than 110,000 GEs, which is negligible compared to the several hundred million gates required for a 16 MB LLC.
\section{Related works on high associativity caches}
\label{sec:related}

\noindent \textbf{Z-Cache}~\cite{zcache} provides higher associativity than traditional set-associative caches with the same number of ways by employing a tag-store walk and a sequence of cache line re-allocations, increasing the number of replacement candidates. However, these candidates are limited to a subset of cache lines, leaving them susceptible to future, faster eviction-set-discovery algorithms. In contrast, Avatar-R eliminates SAEs over a system's lifetime, preventing eviction-set discovery even by advanced future algorithms.
\textbf{HybCache}~\cite{hybcache} is a hybrid cache design combining various caching strategies to enhance performance and security. It uses adaptive techniques to dynamically balance between different cache policies, optimizing for workload demands and minimizing cache contention. However, the authors mention that ``applying HybCache to the LLC or larger caches in general would be expensive (in terms of hardware)". Additionally, HybCache provides security against conflict-based attacks at a performance overhead of up to 5\%, significantly higher than the overhead guaranteed by Avatar-R. HybCache does not provide strong security guarantees against occupancy-based attacks since it does not enforce a strict partitioning, making it weaker than Avatar-P. 
\textbf{Futility Scaling}~\cite{7011401} dynamically adjusts cache associativity within a partition based on performance. Unlike static partitioning or utility-driven approaches, it reduces associativity when the performance gains from additional ways diminish, ensuring more efficient cache utilization by adapting to workload behavior. 
This allows for increased performance, but unlike Avatar-P, it cannot protect against occupancy-based attacks.

 
 
\section{Conclusions}


We present Avatar, a secure, morphable LLC with three dynamic operation modes—\textit{non-secure} (Avatar-N), \textit{randomized secure} (Avatar-R), and \textit{partitioned secure} (Avatar-P). Designed like a conventional set-associative LLC with no decoupling, Avatar ensures easy industry adoption. Avatar-R leverages high associativity with a lightweight Mirage-like design to maintain security against conflict-based attacks while preserving cache capacity. Additionally, it morphs into a partitioned LLC to defend against occupancy-based attacks in Avatar-P. When security is unnecessary, the cache can switch to Avatar-N to operate as a traditional LLC, optimizing performance and power. Avatar-R provides a stronger security guarantee against conflict-based attacks than Mirage--one SAE in a $10^{30}$ years--and Avatar-P mitigates both conflict-based and occupancy-based attacks while outperforming state-of-the-art way-partitioned LLCs. 

\appendix

\begin{figure*}[ht]
    \centering
    \includegraphics[width=0.94\linewidth]{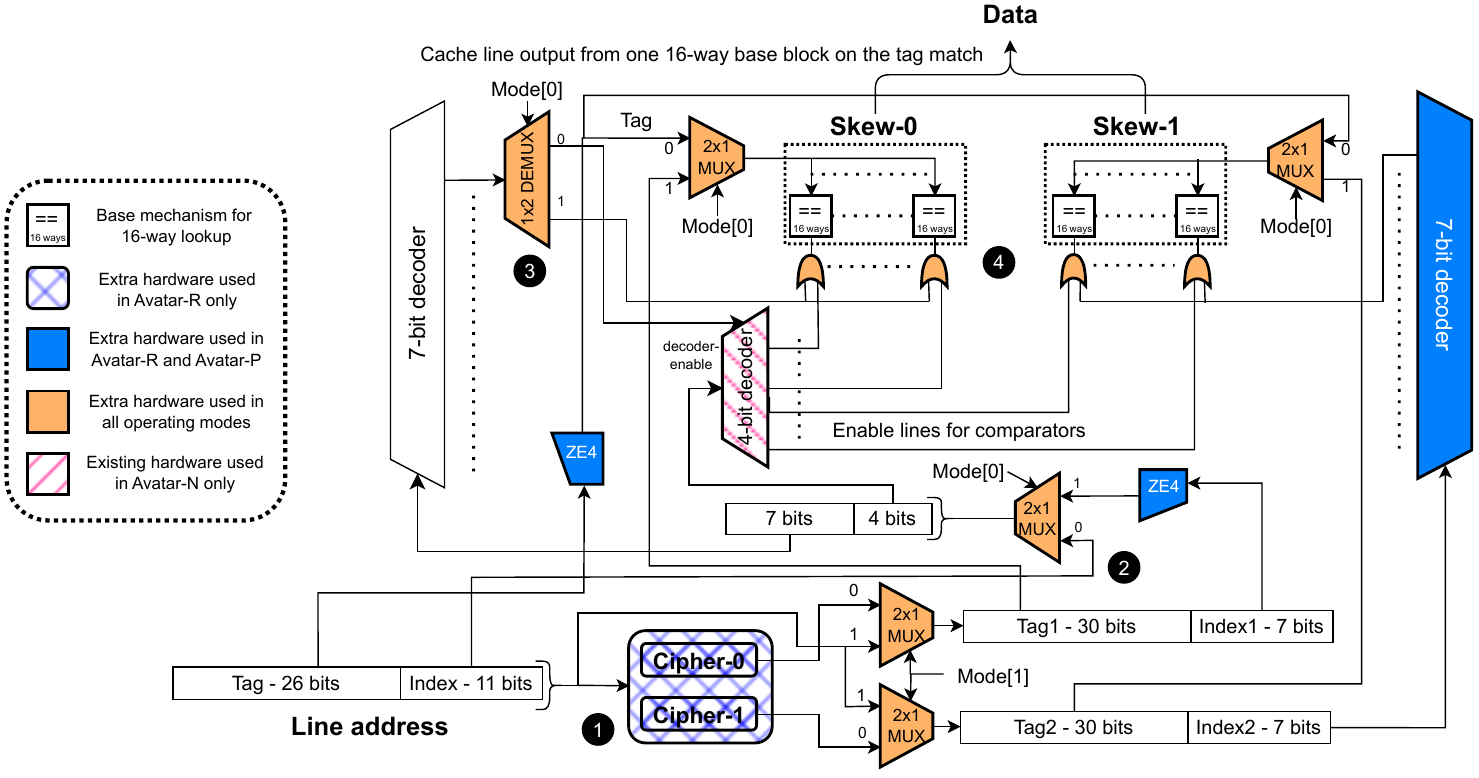}
    \vspace{-0.05 in}
    \caption{Avatar in action. \texttt{Mode} is a 2-bit vector. \texttt{Mode="00"} refers to the Avatar-N, \texttt{Mode="01"} refers to Avatar-R, and \texttt{Mode="11"} refers to Avatar-P. The extra hardware required for morphability has been marked in the figure, along with how the different components are power-gated as per their use.} 
    \label{fig:oasis-circuit}
\end{figure*}

\appendices

\section{Data Availability}

Our work proposes a new cache design that provides strong security guarantees against both conflict-based and occupancy-based attacks. We use a security model based on the bucket-and-balls model to argue about the extremely low probability of getting an SAE in Avatar-R. The performance evaluation for all LLC designs discussed has been done by modeling them in the ChampSim open-source microarchitecture simulator~\cite{champsim}. We extend the existing open-source artifact of Maya~\cite{maya} which already implements some of these secure LLC designs. The power consumption and energy numbers have been obtained using the open-source P-CACTI~\cite{PCACTI} simulator. We plan to publicly release our functional artifact, ensuring full reproducibility of our results.

\section{Avatar LLC in Action}
\label{appendix:A}
In Avatar-R, the line address feeds into two independent ciphers (Cipher-1 and Cipher-2) that generate hashed addresses (\circled{1} in Figure \ref{fig:oasis-circuit}). For Avatar-N and Avatar-P, the ciphers are power-gated, and 11 index bits and seven index bits, respectively, from the original address are used instead of hashed addresses. A 2x1 MUX (\circled{2} in Figure \ref{fig:oasis-circuit}) selects the 11 index bits or the first hash address 7 bits (zero-extended by 4), depending on the mode. Another pair of MUXs chooses between the original and hashed addresses depending on the secure mode. A 7-bit decoder decodes the seven msb bits to enable the appropriate set for tag matching. To differentiate between the address tags for Avatar-R and Avatar-P, we add two 2x1 MUXs before tag-matching. In Avatar-R and Avatar-P, 256 tag ways are matched across two skews, while only 16 ways are matched in Avatar-N. A 1x2 DEMUX (\circled{3} in Figure \ref{fig:oasis-circuit}) enables a 4-bit decoder in Avatar-N to select one of 16 blocks instead of enabling all 16 16-way blocks. In Avatar-R and Avatar-P, all 128 ways in Skew-1 are enabled, with another 7-bit decoder handling Skew-2 using the second hash address's seven bits. To ensure correct functioning in all operation modes, we add OR gates (\circled{4} in Figure \ref{fig:oasis-circuit}), which enable the correct 16-way blocks for tag-matching based on the mode of operation. The tag matches with one of the 16-way blocks and outputs the data from the corresponding cache line. Figure \ref{fig:oasis-circuit} highlights the hardware components activated in each Avatar operation mode. During mode transitions, unused components are power-gated to reduce energy consumption.

\section{Probability of an SAE with Avatar-R.}
\label{appendix:B}
To estimate the probability of an SAE for Avatar, we use the bucket-and-balls model as described in Mirage. The buckets represent cache sets, the balls denote tag entries, and a ball throw represents a fill. The buckets are initialized with as many balls as the cache capacity. This ensures that we model the best-case scenario for the attacker. On a ball throw, two random buckets are chosen, and the ball is inserted into the bucket that is less occupied. If both buckets are at full capacity, we will get a bucket spill representing an SAE. For our analysis, a spill-free scenario is modeled, where the buckets have unlimited capacity. Following the methodology outlined by Mirage, we end up with Equation \ref{eq:analytical_1}. Note that the numerator has 121 ways per skew because 7 out of the 128 ways per skew are invalidated in Avatar-R.

\vspace{-1em}
\begin{equation}
    \scalebox{0.79}{$ \displaystyle \hspace{-1em}\Pr(n \! = \! N \! + \! 1) \! = \! \frac{121}{N \! + \! 1} \! \times \! \Big( \Pr(n \! = \! N)^2 + 2 \! \times \! \Pr(n \! = \! N) \! \times \! \Pr(n \! > \! N) \Big)$}
    \label{eq:analytical_1}
\end{equation}

\vspace{0.5em}
where $\Pr(n \! = \! N \! + \! 1)$ represents the probability of a bucket having $N \! + \! 1$ balls and $\Pr(n \! > \! N \! + \! 1)$ represents the probability of a bucket having more than $N \! + \! 1$ balls. As we increase $N$, $\Pr(n \! = \! N) \! \rightarrow \! 0$ and therefore $\Pr(n \! > \! N) \! \ll \! \Pr(n \! = \! N)$. Using this approximation, Equation \ref{eq:analytical_1} can be simplified to Equation \ref{eq:approx} for larger values of $n$. Similar to the security analysis for Mirage, we only use this approximation once $\Pr(n \! = \! N)$ becomes smaller than 0.01.

\vspace{-0.5em}
\begin{equation}
    \Pr(n \! = \! N \! + \! 1) = \frac{121}{N \! + \! 1} \! \times \! \Pr(n \! = \! N)^2
    \label{eq:approx}
    \vspace{0.5em}
\end{equation}

\begin{figure}[h]
    \centering
    \includegraphics[width=\linewidth]{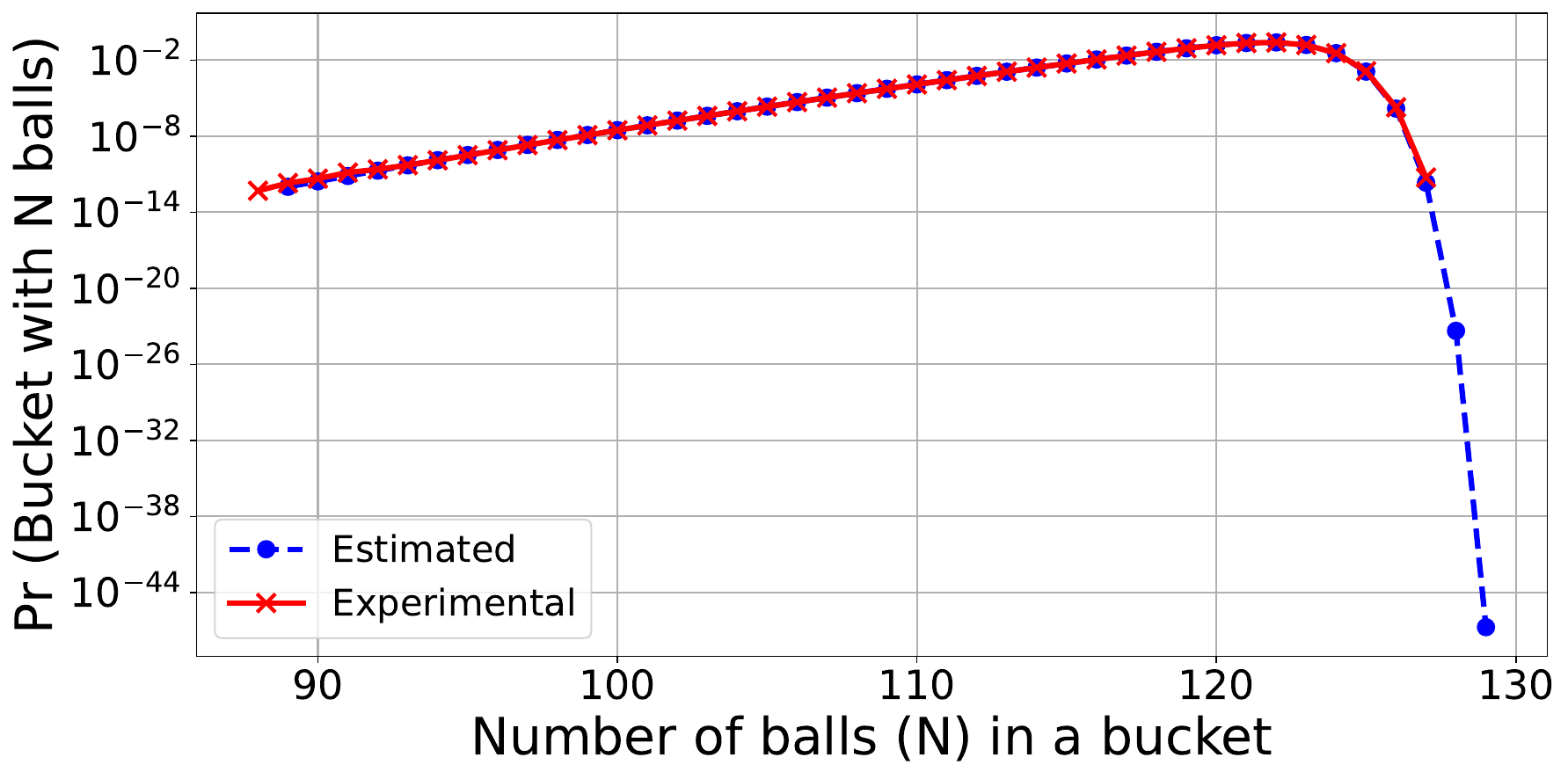}
    \caption{Probability of a bucket having N balls ($\Pr(n = N)$) - experimental and estimated using the analytical model.}
    \label{fig:analytical}
\end{figure}

\noindent \textbf{Frequency of spills.}
We simulate the bucket-and-ball model for one trillion iterations and obtain the probability $\Pr_{exp}(n \! = \! 88) \! \approx \! 3.76 \times 10^{-13}$. Per our simulations, there is no bucket with a ball count of less than 92. Using this value in Equation \ref{eq:analytical_1}, we recursively calculate $\Pr_{est}(n \! = \! N \! + \! 1)$ for $N \! \in \! [89,124]$. Once the probability becomes less than 0.01 ($N \! \in \! [125,128]$), we use Equation \ref{eq:approx}.
If we consider a cache design with W ways per skew, then the probability of an SAE (i.e. a bucket spill) will be given by $\Pr(n \! = \! W \! + \! 1)$. The spill probability follows a double-exponential reduction. For $W \! = \! 126,127,128$, an SAE occurs every $10^{11}$, $10^{23}$, and $10^{46}$ line installs, respectively. Thus, the Avatar LLC design with 128 ways per skew in randomized secure mode has a frequency of one SAE in over $5 \! \cdot \! 10^{46}$ line installs or $10^{30}$ years, which is even more robust than Mirage.

\bibliographystyle{IEEEtran}
\bibliography{references}

\end{document}